\documentclass[aps,twocolumn,showpacs,preprintnumbers,nofootinbib,prd,
superscriptaddress,groupedaddress,10pt]{revtex4-1}
\usepackage[utf8]{inputenc}
\usepackage{graphicx,amssymb,amsmath,amsthm,amsfonts,epsfig,times,natbib}
\usepackage[usenames,dvipsnames]{color}
\usepackage{epstopdf}
\usepackage{aas_macros}
\usepackage{tensor}
\usepackage{mathtools}
\usepackage{mathrsfs}
\usepackage{amsbsy}
\usepackage{bm,url}
\usepackage[linktocpage]{hyperref}
\usepackage[scaled]{beramono}
\usepackage[T1]{fontenc}
\usepackage{mathrsfs}
\usepackage{ulem}
\usepackage{soul}
\usepackage[usenames]{color}

\newcommand{\be}{\begin{equation}}
\newcommand{\ee}{\end{equation}}
\newcommand{\nn}{\nonumber}

\definecolor{oxfordblue}{rgb}{0.0, 0.13, 0.28}
\definecolor{burgundy}{rgb}{0.5, 0.0, 0.13}
\definecolor{darkolivegreen}{rgb}{0.33, 0.42, 0.18}
\definecolor{darkblue}{rgb}{0,0,0.5}
\definecolor{richcarmine}{rgb}{0.84, 0.0, 0.25}
\definecolor{darkblue}{rgb}{0,0,0.5}
\definecolor{venetianred}{rgb}{0.78, 0.03, 0.08}
\definecolor{skobeloff}{rgb}{0.0, 0.48, 0.45}
\hypersetup{colorlinks=true, citecolor=darkblue, linkcolor=darkblue,
urlcolor = darkblue}

\def\non{\nonumber}
\def\nn{\nonumber}

\newcommand{\ben}{\begin{enumerate}}
\newcommand{\een}{\end{enumerate}}

\def\be{\begin{equation}}
\def\ee{\end{equation}}
\def\nn{\nonumber}
\newcommand{\beq}{\begin{eqnarray}}
\newcommand{\eeq}{\end{eqnarray}}
\newcommand{\ba}{\begin{array}}
\newcommand{\ea}{\end{array}}

\newcommand{\comment}[1]{{{{#1}} }}

\begin{document}
%%%%%%%%%%%%%%%%%%%%%%%%%%%%%

\title{Ergoregion instability of exotic compact objects: electromagnetic and
gravitational perturbations and the role of absorption}

\author{
Elisa Maggio$^{1,2,4}$,
Vitor Cardoso$^{2,3}$,
Sam R. Dolan$^{1}$,
Paolo Pani$^{4}$
}
\affiliation{${^1}$ Consortium for Fundamental Physics,
School of Mathematics and Statistics, University of Sheffield, Hicks Building,
Hounsfield Road, Sheffield S3 7RH, United Kingdom}
\affiliation{${^2}$ Centro de Astrof\'{\i}sica e Gravita\c c\~ao  - CENTRA, Departamento de F\'{\i}sica, Instituto Superior T\'ecnico - IST, Universidade de Lisboa - UL, Av. Rovisco Pais 1, 1049-001 Lisboa, Portugal}
\affiliation{${^3}$ Perimeter Institute for Theoretical Physics, 31 Caroline Street North
Waterloo, Ontario N2L 2Y5, Canada}
\affiliation{${^4}$ Dipartimento di Fisica, ``Sapienza'' Universit\`a di Roma \& Sezione INFN Roma1, Piazzale
Aldo Moro 5, 00185, Roma, Italy}

\begin{abstract}
Spinning horizonless compact objects may be unstable against an ``ergoregion
instability''. We investigate this mechanism for electromagnetic
perturbations of ultracompact Kerr-like objects with a reflecting surface,
extending previous (numerical and analytical) work limited to the scalar
case. We derive an analytical result for the frequency and the instability time
scale of unstable modes which is valid at small frequencies.
We argue that our analysis can be directly extended to gravitational
perturbations of exotic compact objects in the black-hole limit.
The instability for electromagnetic and gravitational
perturbations is generically stronger than in the scalar case and it requires
larger absorption to be quenched. We argue that exotic compact
objects with spin $\chi\lesssim 0.7$ ($\chi\lesssim 0.9$) should have an
absorption coefficient of at least $0.3\%$ ($6\%$) to remain linearly stable, and that an 
absorption coefficient of at least $\approx60\%$ would quench the
instability for any spin.
We also show that --~in the static limit~-- the scalar, electromagnetic, and
gravitatonal perturbations of the Kerr metric are related to one another
through Darboux transformations. Finally, correcting previous results, we give the 
transformations that bring the Teukolsky equation in a form described by a real 
potential also in the gravitational case.
\end{abstract}

\maketitle

%%%%%%%%%%%%%%%%%%%%%%%%%%%%%%%%%%%%%%%%%%%%%%%%%%%%
\section{Introduction}
%%%%%%%%%%%%%%%%%%%%%%%%%%%%%%%%%%%%%%%%%%%%%%%%%%%%
Exotic compact objects~(ECOs) are under intense scrutiny as probes of near-horizon quantum
structures~\cite{Cardoso:2016oxy,Cardoso:2017cqb,Cardoso:2017njb}, as models for
exotic states of matter in ultracompact stars~\cite{Pani:2018flj}, and even as exotic 
gravitational-wave~(GW) sources that might coexist in the universe along with black 
holes~(BHs) and neutron stars \comment{as in certain dark-matter scenarios} (see 
Refs.~\cite{Cardoso:2017cqb,Cardoso:2017njb,Barack:2018yly} for recent
overviews).

The phenomenology of ECOs depends strongly on their compactness (or,
equivalently, on the gravitational redshift $z$ at their surface). Objects with
$z\sim{\cal O}(1)$ (e.g., boson stars~\cite{Ruffini:1969qy,Liebling:2012fv})
have properties similar to those of neutron stars and display ${\cal O}(1)$
corrections in most observables (e.g., multipole
moments~\cite{Ryan:1996nk,Pani:2015tga,Uchikata:2015yma,Uchikata:2016qku},
geodesic motion~\cite{Cunha:2015yba}, quasinormal mode~(QNM)
ringing~\cite{Chirenti:2007mk,Pani:2009ss,Macedo:2013jja,Chirenti:2016hzd},
tidal
Love numbers~\cite{Wade:2013hoa,Cardoso:2016rao,Sennett:2017etc}, etc.) relative to
BHs.

A different class of ECOs (dubbed \emph{ClePhOs}
in the terminology introduced in Refs.~\cite{Cardoso:2017cqb,Cardoso:2017njb}) is instead
associated to modifications of the Kerr metric only
very close to the horizon, as in some quantum-gravity
scenarios~\cite{Mazur:2001fv,Mathur:2005zp,Skenderis:2008qn,Almheiri:2012rt,
Giddings:2014ova,Holdom:2016nek,Brustein:2017kcj,Barcelo:2017lnx}.
Here then $z\sim O(10^{20})$ or larger for objects with mass $M\sim 10\,
M_\odot$ or higher. It is
extremely challenging --~if not impossible~\cite{Abramowicz:2002vt,Cardoso:2017cqb,Cardoso:2017njb}~-- to
rule out or detect these objects through electromagnetic observations, since
their geodesic structure is almost identical to that of a BH.
On the other hand, GWs emitted by these objects in different scenarios
carry unique information on their properties. This includes: (i)~GW echoes
in the postmerger ringdown phase of a binary
coalescence~\cite{Cardoso:2016rao,Cardoso:2016oxy} (see also
Ref.~\cite{Ferrari:2000sr} for an earlier study, and
Refs.~\cite{Abedi:2016hgu,Conklin:2017lwb,Ashton:2016xff,Abedi:2017isz,
Westerweck:2017hus,Abedi:2018pst} for a debate on the evidence of this effect in
aLIGO data); (ii) a (logarithmically-small)
tidal deformability that affects the late
inspiral~\cite{Cardoso:2017cfl,Maselli:2017cmm}; (iii) the absence
of tidal heating for ECOs as compared to BHs~\cite{Maselli:2017cmm}; (iv) their
different spin-induced quadrupole
moment~\cite{Pani:2015tga,Uchikata:2015yma,Uchikata:2016qku,Krishnendu:2017shb};
and (v) the stochastic GW background from a population of
ECOs~\cite{Du:2018cmp,Barausse:2018vdb,Fan:2017cfw}.

In parallel with developing detection strategies for these signatures,
it is also important to assess the viability of ECOs and, in particular, their
stability and their formation channels.
Spinning ECOs are potentially unstable, due to the so-called \emph{ergoregion
instability}~\cite{1978CMaPh..63..243F} (for a review, see
Ref.~\cite{Brito:2015oca}). The latter is an instability that develops in any
asymptotically flat spacetime featuring an ergoregion but without an event
horizon: since physical negative-energy states can exist inside the ergoregion
--~a key ingredient in Penrose's process~\cite{Penrose:1969}~-- it is
energetically favorable to cascade toward even more negative states. The only
way to prevent such process from developing is by absorbing the
negative-energy states. Kerr BHs can absorb radiation very efficiently and are
indeed stable even if they have an ergoregion. On the other hand, compact
horizonless geometries are generically unstable in the absence of dissipation mechanisms.

The ergoregion instability has been studied for various
models~\cite{1978CMaPh..63..243F,Vilenkin:1978uc,1978RSPSA.364..211C,
1996MNRAS.282..580Y,Kokkotas:2002sf,Brito:2015oca,Cardoso:2007az,Cardoso:2008kj,
Pani:2010jz}. The time scale of the instability depends strongly on the spin
and on the compactness of the object~\cite{Cardoso:2007az,Chirenti:2008pf,Brito:2015oca}. In particular, the
instability exists only for those objects which are compact enough to possess a
photon sphere~\cite{Cardoso:2014sna,Brito:2015oca}. The latter is
naturally present in all models of ECOs that modify the BH geometry only at the
horizon scale, for example by invoking an effective surface located at some
Planck distance from the would-be
horizon~\cite{Cardoso:2017cqb,Cardoso:2017njb}.

The ergoregion instability of Planck-inspired ECOs has been recently studied
in Ref.~\cite{Maggio:2017ivp} for scalar-field perturbations. There, it was
 shown that, if the ECO interior does not absorb any radiation, the
instability time scale is short enough to have a crucial impact on the dynamics
of the object. On the other hand, partial absorption (i.e.
reflectivity at the object's surface smaller than unity) may quench the
instability completely, just like in the BH case. Because the dissipation
of energy in compact objects made of known matter is small, the instability imposes severe constraints
on some particular models of horizonless objects~\cite{Barausse:2018vdb}.

In this paper we extend the analysis of Ref.~\cite{Maggio:2017ivp} to electromagnetic and gravitational perturbations. Since the ergoregion instability is intimately linked to
superradiance~\cite{Brito:2015oca,Vicente:2018mxl}, and since superradiance is
enhanced by field spin (at least for rapidly-spinning objects), one expects
that the instability gets stronger for gravitational perturbations.
Below we quantify this expectation --~both by solving the full linear problem
numerically and by computing the spectrum of unstable modes analytically in the
small-frequency limit~-- and we discuss the implications of our results for
current observational constraints on ECOs.

Any instability is of course also related to the boundary conditions of the problem. While formulating physical  boundary conditions for electromagnetic and gravitational fluctuations, we uncovered a curious relation between all sets of perturbations
in the static limit: scalar, electromagnetic and gravitational perturbations of the Kerr metric are all related to one another
through Darboux transformations. To the best of our knowledge this interesting
property of BH perturbations in general relativity has not been
reported before.
Through this work, we use $G=c=1$ units.

%%%%%%%%%%%%%%%%%%%%%%%%%%%%%%%%%%%%%%%%%%%%%%%%%%%%
\section{Setup}
%%%%%%%%%%%%%%%%%%%%%%%%%%%%%%%%%%%%%%%%%%%%%%%%%%%%

%%%%%%%%%%%%%%%%%%%%%%%%%%%%%%
\subsection{Kerr-like ECO model}
%%%%%%%%%%%%%%%%%%%%%%%%%%%%%%

Our setup and methods follow Refs.~\cite{Maggio:2017ivp,Barausse:2018vdb}.
We consider a geometry described by the Kerr metric\footnote{
The vacuum region outside a spinning object is not necessarily described by the Kerr geometry,
due to the absence of an analog to the Birkhoff's theorem in axisymmetry.
This implies that the multipolar structure of a spinning ECO might be
generically different from that of a Kerr BH. However, in those models that
admit a smooth BH limit, there are
indications that all multipole moments of the external spacetime approach those
of a Kerr BH as $\epsilon\to0$~\cite{Pani:2015tga,Uchikata:2015yma,
Uchikata:2016qku,Yagi:2015hda,Yagi:2015upa,Posada-Aguirre:2016qpz}.
In fact, for $z\to\infty$, it is
natural
to expect that the exterior spacetime is extremely close to Kerr, unless some
discontinuity occurs in the BH limit. See
Refs.~\cite{Mark:2017dnq,Maggio:2017ivp,Barausse:2018vdb,Abedi:2016hgu,
Wang:2018gin, Burgess:2018pmm} for other work discussing the same model.
}
when $r>r_0$ and, at $r=r_0$, we assume the presence of a membrane with some
reflective properties. Different models of ECOs are characterized by
different properties of the membrane at $r=r_0$, in particular by a
(possibly) frequency-dependent
reflectivity~\cite{Mark:2017dnq,Maggio:2017ivp}. In
Boyer-Lindquist coordinates,
the line element at $r>r_0$ reads
\begin{eqnarray}
ds^2&&=-\left(1-\frac{2Mr}{\Sigma}\right)dt^2+\frac{\Sigma}{\Delta}dr^2-\frac{
4Mr}{\Sigma}a\sin^2\theta d\phi dt   \nn \\
&+&{\Sigma}d\theta^2+
\left[(r^2+a^2)\sin^2\theta +\frac{2Mr}{\Sigma}a^2\sin^4\theta
\right]d\phi^2\,,\label{Kerr}
\end{eqnarray}
where $\Sigma=r^2+a^2\cos^2\theta$ and $\Delta=r^2+a^2-2M r$, with $M$ and
$J:=aM$ the total mass and spin of the object.

Motivated by models of microscopic corrections at the horizon scale, in the
following we shall focus on the case
%%%
\begin{equation}
 r_0 = r_+(1+\epsilon) \qquad 0<\epsilon\ll 1\,,  \label{epsilon-def}
\end{equation}
where $r_+=M+\sqrt{M^2-a^2}$ is the location of the would-be horizon. Although
the above parametrization requires $a\leq M$, the latter condition is not
strictly necessary~\cite{Maggio:2017ivp}; the case of so-called ``superspinars'' (when $a>M$~\cite{Gimon:2007ur,Cardoso:2008kj,Pani:2010jz,Maggio:2017ivp}) will be discussed elsewhere.
Note that $\epsilon$ is related to the compactness of the object and to the
gravitational redshift at the surface, namely
$M/r_0 \approx M/r_+ (1-\epsilon)$ and $z \approx \epsilon^{-1/2}(r_+/M-1)^{-1/2}$.
%%%
If $r_0\sim r_+ +l_P$ (where $l_P$ is the Planck length, as suggested by some
quantum-gravity inspired
models~\cite{Cardoso:2016oxy,Cardoso:2017cqb,Cardoso:2017njb}), then
$\epsilon\sim 10^{-40}$ for a non-spinning object with $M \sim 50 \,
M_{\odot}$.

%%%%%%%%%%%%%%%%%%%%%%%%%%%%%%%%%
\subsection{Linear perturbations}
%%%%%%%%%%%%%%%%%%%%%%%%%%%%%%%%%

Scalar, electromagnetic, and gravitational perturbations in the exterior Kerr
geometry are described in terms of Teukolsky's master
equations~\cite{Teukolsky:1972my,Teukolsky:1973ha,Teukolsky:1974yv}
\begin{eqnarray}
&&\Delta^{-s} \frac{d}{dr}\left(\Delta^{s+1} \frac{d_{s}R_{lm}}{dr}\right)\nonumber\\
&+& \left[\frac{K^{2}-2 i s (r-M) K}{\Delta}+4 i s \omega r -\lambda\right] \ _{s}R_{l m}=0\,,\label{wave_eq} \quad \\
&&\left[\left(1-x^2\right)~_{s}S_{l m,x}\right]_{,x}+ \bigg[(a\omega x)^2-2a\omega s x + s \nonumber\\
&+&~_{s}A_{lm}-\frac{(m+sx)^2}{1-x^2}\bigg]~_{s}S_{l m}=0\,, \label{angular}
\end{eqnarray}
where $~_{s}S_{l m}(\theta)e^{im\phi}$ are spin-weighted spheroidal
harmonics, $x\equiv\cos\theta$, $K=(r^2+a^2)\omega-am$, and the separation
constants
$\lambda$ and $~_{s}A_{l m}$ are related by $\lambda \equiv  ~_{s}A_{l
m}+a^2\omega^2-2am\omega$.
When $a=0$, the angular eigenvalues are $\lambda=(l-s)(l+s+1)$, whereas for $a\neq0$ they can be computed numerically or with approximated analytical expansions (see Sec.~\ref{sec:numproc}) below). 

It is convenient to make a change of variables by introducing
Detweiler's function~\cite{1977RSPSA.352..381D}
\begin{equation}
 _{s}X_{lm} = \Delta^{s/2} \left(r^2+a^2\right)^{1/2} \left[\alpha \
_{s}R_{lm}+\beta \Delta^{s+1} \frac{d_{s}R_{lm}}{dr}\right]\,,\label{DetweilerX}
\end{equation}
where $\alpha$ and $\beta$ are certain radial functions. Introducing the
tortoise coordinate $r_*$, defined such that
$dr_*/dr=(r^2+a^2)/\Delta$, the master equation~\eqref{wave_eq} becomes
%%%
\begin{equation}
 \frac{d^2_{s}X_{lm}}{dr_*^2}- V(r,\omega) \,_{s}X_{lm}=0\,, \label{final}
\end{equation}
%%%
where the effective potential is
%%%
\begin{equation}
 V(r,\omega)=\frac{U\Delta }{(r^2+a^2)^2}+G^2+\frac{dG}{dr_*}\,, \label{pot_detweiler}
\end{equation}
%%%
and
\beq
G &=& \frac{s(r-M)}{r^2+a^2}+\frac{r \Delta}{(r^2+a^2)^2} \,, \\
U &=& V_S+\frac{2\alpha' + (\beta' \Delta^{s+1})'}{\beta \Delta^s} \,, \\
V_S &=& -\frac{1}{\Delta}\left[K^2-is\Delta'K+\Delta(2isK'-\lambda)\right] \,.
\eeq
The prime denotes a derivative with respect
to $r$ and the functions $\alpha$ and
$\beta$ can be chosen such that the resulting potential is purely
real (see Ref.~\cite{1977RSPSA.352..381D} for the definitions of $\alpha$ and 
$\beta$ in the electromagnetic case and Appendix~\ref{app:transformation} for the 
gravitational case\footnote{Ref.~\cite{1977RSPSA.352..381D} has some mistakes in 
the definitions of the radial functions $\alpha$ and $\beta$ in the gravitational case, 
which we correct in Appendix~\ref{app:transformation}.}). 
In the following we define $R_s \equiv~_{s}R_{lm}$, $X_s \equiv~_{s}X_{lm}$ and omit the $l$, $m$ subscripts for brevity.

%%%%%%%%%%%%%%%%%%%%%%%%%%%%%%%%
\subsection{Boundary conditions}
%%%%%%%%%%%%%%%%%%%%%%%%%%%%%%%%

By imposing boundary conditions at infinity and at the surface of the ECO, Eq.~\eqref{final} defines an eigenvalue problem whose eigenvalues, $\omega=\omega_R+i\,\omega_I$, are the QNMs of the system. In our convention a stable mode corresponds to $\omega_I<0$, whereas an unstable mode corresponds to $\omega_I>0$ with instability timescale $\tau := 1/\omega_I$.

In order to derive the QNM spectrum, we impose outgoing boundary conditions at
infinity~\cite{Teukolsky:1974yv}
\be
X_s \sim e^{i\omega r_*} \qquad r \rightarrow \infty \,. \label{BCinf}
\ee
At $r=r_0$ there is a superposition of ingoing and outgoing
waves. In general, the boundary condition depends on the properties of the
membrane of the ECO. In the following we will mainly focus on the analysis of a perfectly reflecting surface -- a discussion about the role of partial absorption by the object is presented in Sec.\ref{sec:absorption}. In the case of electromagnetic perturbations, we consider a perfect conductor in which the electric and
magnetic fields satisfy $E_\theta (r_0) = E_\phi (r_0) = 0$ and $B_r (r_0) = 0$. By
writing the previous conditions in terms of the three complex scalars of the electromagnetic
field in the Newman-Penrose formalism~\cite{1977MPCPS..81..149K}, we obtain
that the following boundary conditions on Teukolsky's function~\cite{Brito:2015oca}
%%%%%
\begin{equation}
\partial_{r} R_{-1}  = \Bigg[ \frac{iK}{\Delta} - \frac{i}{2 K} \Big(\lambda
\pm B + 2 i \omega r \Big) \Bigg] R_{-1} \,, \label{BCr0}
\end{equation}
%%%
% 
where\footnote{We notice that Ref.~\cite{Brito:2015oca} has a
typo in the definition of $B$.} $B = \sqrt{\lambda^2 +4ma\omega-4a^2 \omega^2}$, the plus
and minus signs refer to polar and axial perturbations, respectively.
Note that the above boundary conditions define a perfectly conducting object, since the outgoing energy flux is equal to the incoming flux at the surface. In Sec.~\ref{subsec:analyticem} we will show that the boundary conditions \eqref{BCr0} are equivalent to the following boundary  conditions on Dirichlet's function, when $\epsilon \ll 1$,
%%%%
\begin{equation}
 \left\{\begin{array}{ll}
    X_{-1}(r_0)=0 & \qquad{\rm axial} \\
    dX_{-1}(r_0)/dr_*=0 &\qquad {\rm polar} \\
   \end{array}\right.\,. \label{BCsXem}
\end{equation}
%%%%
%
Motivated by a "bounce-and-amplify" argument (see Sec.~\ref{sec:anagrav} below), we shall 
extend this result to gravitational perturbations of a perfectly reflecting ECO, in 
which case we impose
%%%%
\begin{equation} 
 \left\{\begin{array}{ll}
    X_{-2}(r_0)=0 & \qquad{\rm axial} \\
    dX_{-2}(r_0)/dr_*=0 &\qquad {\rm polar} \\
   \end{array}\right.\,. \label{BCsXgrav}
\end{equation}
%%%%

%%%%%%%%%%%%%%%%%%%%%%%%%%%%%%%%%
\subsection{Numerical procedure} \label{sec:numproc}
%%%%%%%%%%%%%%%%%%%%%%%%%%%%%%%%%

Equation~\eqref{final} with boundary conditions~\eqref{BCinf} and~\eqref{BCr0} [or \eqref{BCsXem}], \eqref{BCsXgrav}]
can be solved numerically through a direct integration shooting
method~\cite{Pani:2013pma}. Starting with a high-order series expansion at
large distances, we integrate Eq.~\eqref{final} [or, equivalently,
Eq.~\eqref{wave_eq}] from infinity to $r=r_0$; we repeat the integration for
different values of $\omega$ until the desired boundary condition at $r_0$ is
satisfied.

The QNMs of the system depend on two continuous dimensionless parameters: the
spin $\chi=a/M$ and the parameter $\epsilon$, defined in Eq.~(\ref{epsilon-def}), that is related to the ECO
compactness and to the redshift at the ECO surface. Furthermore,
the QNMs depend on three integer numbers: the angular number $l\geq0$, the
azimuthal number $m$ (such that $|m|\leq l$), and the overtone number $n\geq0$.
We shall focus on the fundamental
modes ($n=0$)with $l=m=1$ for electromagnetic perturbations and $l=m=2$ for gravitational perturbations which, in the unstable case, correspond to the modes with the
largest imaginary part, and thus the shortest instability time scale.
In our numerical results, we also make use of the symmetry~\cite{Leaver:1985ax}
\begin{equation}
 m\to-m\,,\quad\omega_R\to-\omega_R\,,\quad _{s}A_{lm} \to \
_{s}A_{l-m}^*\,.\label{symmetry}
\end{equation}
The latter guarantees that, without loss of generality, we can focus on modes
with $m\geq0$ only.

Finally, the angular eigenvalues can be computed numerically using continued fractions~\cite{Berti:2005gp}.
For $a\omega\ll1$, $~_{s}A_{lm}$ can also be expanded analytically as $~_{s}A_{lm}=\sum_{n=0}f^{(n)}_{slm}(a\omega)^{n}$,
where $f^{(n)}_{slm}$ are known expansion coefficients~\cite{Berti:2005gp}. The above 
expression provides an 
excellent approximation whenever $|a\omega|\lesssim1$. In the numerical results 
presented below we have used the full numerical 
expression of $~_{s}A_{lm}$ obtained through continued fractions. We checked that the 
analytical approximation (up to second order) differs from exact eigenvalues of \comment{$\lesssim 
2\%$}  for the electromagnetic modes at high spin and \comment{$\lesssim 4\%$}  for the gravitational 
modes.

%%%%%%%%%%%%%%%%%%%%%%%%%%%%%%%%%%%%%%%%%%%%%%%%%%%%%%%%%%%%%%%%%%%%
\section{ECO instabilities for electromagnetic perturbations}
%%%%%%%%%%%%%%%%%%%%%%%%%%%%%%%%%%%%%%%%%%%%%%%%%%%%%%%%%%%%%%%%%%%%

%%%%%%%%%%%%%%%%%%%%%%%%%%%%%%%%%%%%%%%%%%%%%%%%%%%%%%%%%%%%%%%%%%%%%%%%%%%%%%%%%%%%%%%%%%%%%
\subsection{Analytical results: Extension of Vilenkin's calculation to electromagnetic
perturbations}
%%%%%%%%%%%%%%%%%%%%%%%%%%%%%%%%%%%%%%%%%%%%%%%%%%%%%%%%%%%%%%%%%%%%%%%%%%%%%%%%%%%%%%%%%%%%%
\label{subsec:analyticem}

Here we extend Vilenkin's analytical computation~\cite{Vilenkin:1978uc} of
scalar perturbations in the background of a perfectly-reflecting Kerr-like object to the
electromagnetic case. We use Detweiler's
transformation~\eqref{DetweilerX} and introduce standard `in' and `up' modes,
denoted $X^+_s$ and $X^-_s$, respectively, with asymptotic behavior
\begin{subequations}
\begin{align}
X^{+}_s &\sim
\begin{cases}
B_+ e^{-i \tilde{\omega} r_{*}}  &r_{*} \rightarrow -\infty\,,  \\
e^{-i \omega r_{*}}+A_+ e^{+i \omega r_{*}} \ \ \ &r_{*} \rightarrow \infty\,,
\end{cases}
 \label{asymp_infty} \\
X^{-}_s &\sim
\begin{cases}
e^{+i \tilde{\omega} r_{*}}+A_- e^{-i \tilde{\omega} r_{*}} \ \ \ & r_{*}
\rightarrow -\infty\,, \\
B_- e^{+i \omega r_{*}} &r_{*} \rightarrow \infty\,,
\end{cases}
 \label{asymp_sol}
 \end{align}
\end{subequations}
where $\tilde{\omega} = \omega - m \Omega$ and $\Omega = a/(2Mr_{+})$ is the
angular velocity at the horizon. Since the effective potential in Eq.~\eqref{final} is real, $X^\pm_s$ and their complex conjugates $X^{\pm\ast}_s$ are independent solutions to the same equation which satisfy complex conjugated boundary conditions. Via the Wronskian relationships, the coefficients $A_\pm$ and $B_\pm$ satisfy the
relations~\cite{PhysRevD.71.124016}
\begin{subequations}
\beq
1-|A_+|^2 &=& \left(\tilde{\omega}/\omega\right) |B_+|^2\,, \label{wronskian1}\\
1-|A_-|^2 &=& \left(\omega/\tilde{\omega}\right) |B_-|^2\,, \label{wronskian2}
\eeq
\end{subequations}
and $\tilde{\omega} B_+ = \omega B_-$, from which it follows that $|A_-|=|A_+|$.

We focus on the solution with asymptotics (\ref{asymp_sol}) and we impose the
boundary conditions~\eqref{BCr0} at the surface, i.e., we assume that the latter
is a perfect conductor.
Near the surface, the function $R_{-1}$ defined in Eq.~\eqref{BCr0} has the
following asymptotics 
\be
R_{-1}^{-} \sim \mathcal{A} \Delta e^{-i \tilde{\omega} r^*} + \mathcal{B} e^{+i
\tilde{\omega} r^*} \label{R-1} \ \ \ \ \ \ \ \ r^* \rightarrow -\infty\,,
\ee
where $\mathcal{A} = \mathcal{A_\text{0}} + \eta \mathcal{A_{\text{1}}} + ...$
and $\mathcal{B} = \mathcal{B_\text{0}} + \eta \mathcal{B_\text{1}} + ...$,
with $\eta \equiv r-r_+$.
Since $\Delta\sim (r_+-r_-)\eta$ near the surface, in Eq.~\eqref{R-1} we
consider $\mathcal{A} = \mathcal{A_\text{0}}$ and $\mathcal{B} =
\mathcal{B_\text{0}} + \eta \mathcal{B_\text{1}}$.
We then obtain
\beq
\mathcal{B_\text{0}} &=& \frac{-2^{1/2} (r_+^2 + a^2)^{1/2} \tilde{\omega}}{B}\,,\label{B0} \\
\mathcal{A_\text{0}} &=& \frac{-i B}{4 K_+ \mathfrak{R}^*} \
\mathcal{B_\text{0}} \ A_-\,, \label{A0}
\eeq
where $K_+ = K(r_+)$, $\mathfrak{R} = i K_+ + (r_+ - r_-)/2$ \cite{PhysRevD.71.124016}. By inserting
Eq.~\eqref{R-1} in the Teukolsky equation, we find
\be
\mathcal{B_\text{1}} = \left(\frac{i a m}{M(r_+ - r_-)} + \frac{2 \omega r_+ -
i \lambda}{4 M r_+ \tilde{\omega}} \right) \mathcal{B_\text0}\,. \label{B1}
\ee
Equation~\eqref{R-1} with Eqs.~\eqref{B0}, ~\eqref{A0}, ~\eqref{B1} defines the asymptotic expansion of $R_{-1}$ near the horizon at the first order in $\eta$. By inserting Eq.~\eqref{R-1} in the boundary condition \eqref{BCr0},
we get the following expression
\begin{equation}
e^{i \tilde{\omega} r_{*}^{0}} \mp A_- e^{-i \tilde{\omega} r_{*}^{0}} = 0\,,
\label{omega_vil0}
\end{equation}
for the two signs of Eq.~\eqref{BCr0}, respectively, which correspond to polar ($-$) and axial ($+$) modes, and where $r_{*}^{0} = r_*(r_0)$. The above equation takes the same form of Eq.~(11) in Ref.~\cite{Vilenkin:1978uc} for scalar perturbations. Note that Eq.~\eqref{omega_vil0} implies that the perfect-conductor boundary conditions~\eqref{BCr0} correspond to
Dirichlet and Neumann boundary conditions on the function $X^-_{-1}$ for axial and polar modes, respectively (see above Eq.~\eqref{BCsXem} for an explicit expression).
Equation~\eqref{omega_vil0} yields
\begin{equation}
\tilde{\omega} = \frac{1}{2 |r_*^0|} \left( p \pi - \Phi + i \ln|A_-| \right)\,,
\label{omega_vil}
\end{equation}
where $p$ is a positive odd (even) integer for axial (polar) modes and $\Phi$ is the phase of the reflected wave at $r=r_0$.

In order to compute the imaginary part of the mode, we first recall
that $|A_-|=|A_+|$, so we need to derive $|A_+|$. For waves originating at
infinity, the `in' mode $X^{+}_{-1}$ has asymptotics (\ref{asymp_infty}). If we express the latter in
terms of Teukolsky's function $R_{-1}$, we find 
\be
R_{-1}^{+} \sim r^{-1} \mathcal{C} e^{-i \omega r_*} + r \mathcal{D} e^{i \omega
r_*} \ \ \ \ \ \ \ \ r_* \rightarrow \infty\,,
\ee
where the coefficients of incident and reflected waves are $\mathcal{C}=1/(2^{3/2} |\omega|)$ and $\mathcal{D}=\frac{4 \omega^2}{B} \mathcal{C} A_{+}$~\cite{PhysRevD.71.124016}. In the electromagnetic case the amplification factor is defined as
\cite{Brito:2015oca}
\be
Z = \frac{|\mathcal{D}|^2}{|\mathcal{C}|^2} \frac{B^2}{16 \omega^4} - 1\,,
\ee
which implies
\be
Z = |A_{+}|^2  -1 \,. \label{Z}
\ee

In the low-frequency regime, the amplification coefficient for generic
spin-$s$ waves in Kerr metric has been computed by
Starobinsky~\cite{Starobinskij2}
\be
Z = -D_{lm} = 4 Q \beta_{sl}\prod_{k=1}^{l} \left(1 + \frac{4 Q^2}{k^2}\right)
\left[\omega (r_+ - r_-)\right]^{2l+1} \,, \label{amplfactor}
\ee
where $\sqrt{\beta_{sl}}=\frac{(l-s)! (l+s)!}{(2l)! (2l+1)!!}$ and $Q = \frac{r_+^2 + a^2}{r_+ - r_-} (m \Omega - \omega)$.
In our calculations we consider $Z = -{\rm Re}\{D_{lm}\}$ since the QNM frequency is
complex and $\omega_I\ll\omega_R$.

By inserting Eq.~\eqref{amplfactor} into Eq.~\eqref{Z} and recalling that $|A_+|=|A_-|$, we derive the imaginary part of the frequency in Eq.~\eqref{omega_vil}
\be
|A_-|^2 - 1 = - {\rm Re}\{D_{lm}\}\,,
\ee
which is analogous to the Vilenkin relation for scalar perturbations.
Note that $Z>0$ (i.e. $\omega_I>0$) in the superradiant regime
$\omega_R(\omega_R-m\Omega)<0$.
We have therefore shown that electromagnetic unstable modes of a
perfectly-reflecting Kerr-like object can be understood in terms of
waves amplified at the ergoregion and being reflected at the boundary.

In Appendix\comment{~\ref{app:matching}} we derive an analogous result using a matched
asymptotic expansion. In addition, the latter allows us to compute
the phase $\Phi$ in Eq.~\eqref{omega_vil} analytically. To summarize, the
analytical result valid at small frequency reads\footnote{Comparison
between Eq.~\eqref{wRana} and Eq.~\eqref{omega_vil} reveals that the phase
$\Phi$ defined in Eq.~\eqref{omega_vil} is a constant and does not depend on the
spin. This analytical result is in contrast with the phase computed numerically in
Ref.~\cite{Maggio:2017ivp}.
Since the phase is computed from a matched asymptotic expansion,
we believe that the spin dependence of $\Phi$ computed in
Ref.~\cite{Maggio:2017ivp} is due to a different approximation in the
calculation. Indeed, in the region near the surface of the ECO, Eq.~(A3) of
Ref.~\cite{Maggio:2017ivp} neglects more terms proportional to $a \omega$ with
respect to Eq.~\eqref{wave_eq_near_hor}.} 
\comment{
%%%
\begin{eqnarray}
 \omega_R&\sim&
- \frac{\pi}{2|r_*^0|}\left[q+\frac{s(s+1)}{2}\right]+m\Omega \,,
\label{wRana} \\
 %%%
 \omega_I &\sim&
-\frac{\beta_{sl}}{|r_*^0|}\left(\frac{2 M r_+}{r_+-r_-}\right)\left[
\omega_R(r_+-r_-)\right]^{2l+1}(\omega_R-m\Omega)\,, \nn\\\label{wIana}
\end{eqnarray}
%%%
}
where $r_*^0\sim M[1+(1-\chi^2)^{-1/2}]\log\epsilon$ and $q$ is a positive odd (even) 
integer for polar (axial) modes.
The above result is valid for $s=-1$, for $s=0$ (in the latter case $q$ is a 
positive odd (even) integer for Neumann (Dirichlet) boundary conditions on the scalar 
field), \comment{and also for gravitational perturbations when $s=-2$; the latter result will be 
discussed in Sec.~\ref{sec:grav}.}
Furthermore, we note that the hypothesis of low frequency implies that $M\omega_R\ll1$. In 
order to fullfil this condition in the spinning case, it is not sufficient that 
$\log\epsilon\ll1$, but also that $M\Omega\ll1$.
Indeed in the BH limit ($\epsilon\to0$) we obtain $\omega_R\sim m\Omega$ (hence the
frequency is independent on $\epsilon$ as long as $\Omega M\gg\epsilon$) and
$\omega_I \sim (m\Omega)^{2l+1}/\log^2\epsilon$. In this limit,
the above analytical result is strictly valid only when
$\Omega M\ll1$.

On the other hand, the analytical result is always accurate near the critical value 
of the spin such that $\omega_R\approx 0$ and $\omega_I\approx 0$. The above equations 
predict that the instability occurs when
$\omega_R(\omega_R-m\Omega)<0$ which, for $\epsilon\to0$, implies \comment{(for $s=-1$)}
%%%
%
\begin{equation}
 \chi>\chi_{\rm crit} \approx\frac{\pi q}{m |\log\epsilon|}\,.
\label{chicrit}
\end{equation}
%%%
Therefore, the critical value of the spin above which the instability occurs
can be very small as $\epsilon\to0$. However, in the same limit the instability
time scale is $\tau\propto \log^2\epsilon/
(m\Omega)^{2l+1}$~\cite{Maggio:2017ivp}.
In other words, as $\epsilon\to0$ also slowly-spinning ECOs become
unstable. In the same limit their instability time scale can be very
long but still relevant on dynamical scales~\cite{Maggio:2017ivp}.

%
%%%%%%%%%%%%%%%%%%%%%%%%%%%%%%%%%%%%%%%%%%%%%%%%%%%%
\subsection{Numerical results}
\begin{figure*}[th]
    \centering
    \includegraphics[width=0.45\textwidth]{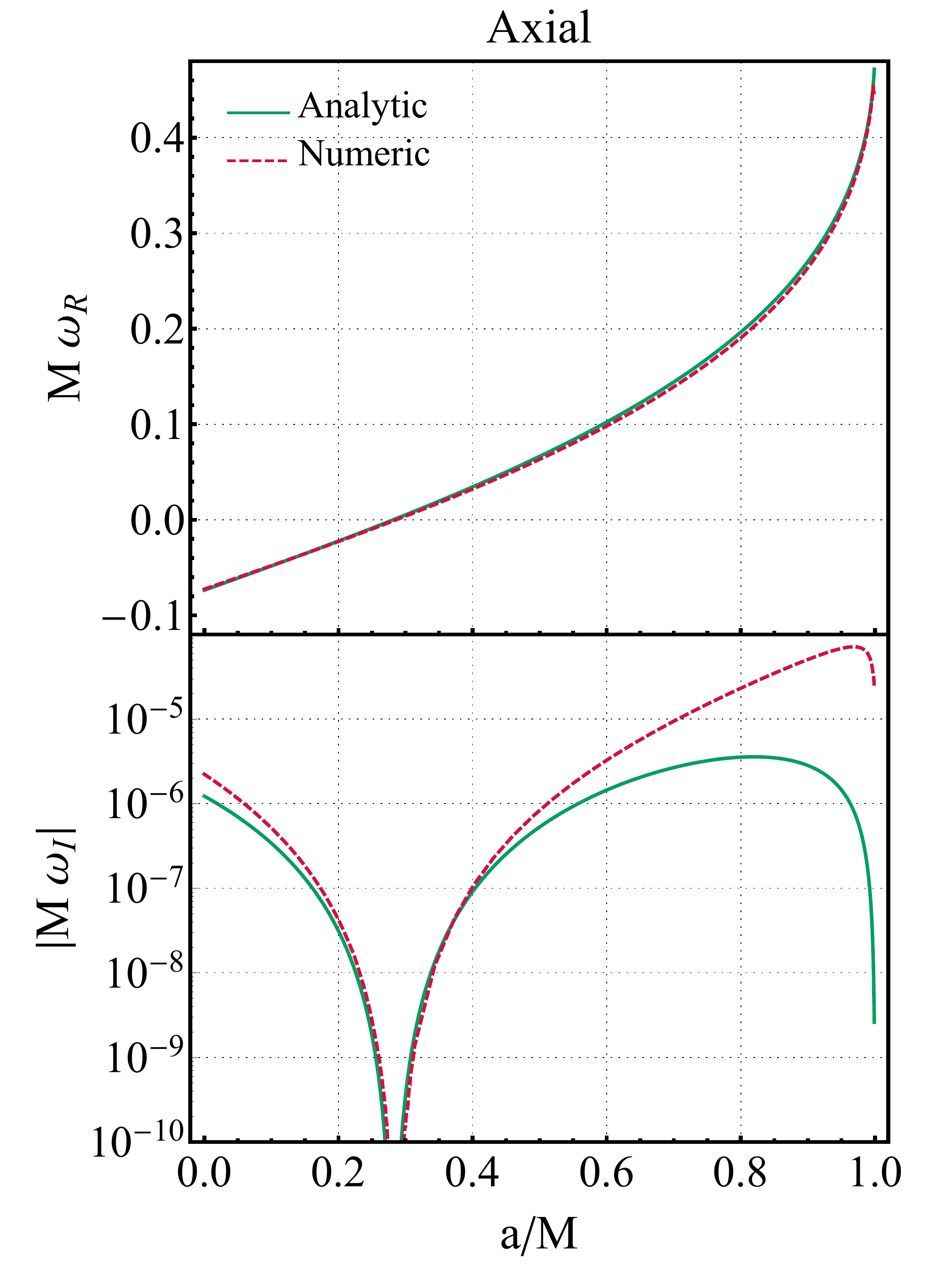}
    \includegraphics[width=0.45\textwidth]{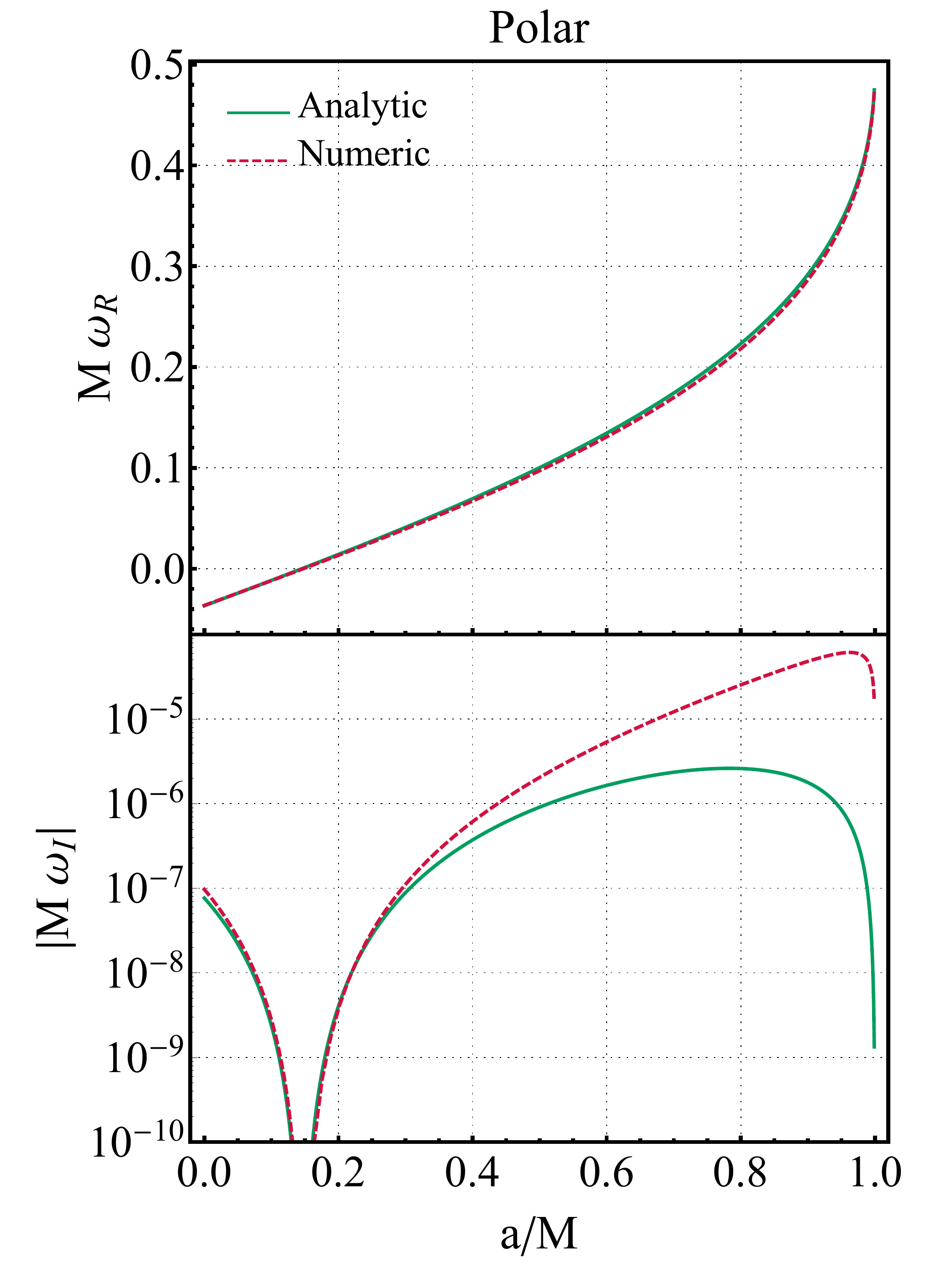}
    \caption{Real (top panels) and imaginary (bottom panels) part of the fundamental
electromagnetic QNM ($l=m=1$, $n=0$) of an ECO as a function of the spin. The left (right) panels refer to axial (polar) modes [corresponding to minus (plus) sign in
Eq.~\eqref{BCr0}]. The surface of the ECO is located at $r_0 = r_+ (1+\epsilon)$ with $\epsilon = 10^{-10}$. The QNMs computed numerically
(dashed curves) are in agreement with the QNMs computed analytically through
Eqs.~\eqref{wRana}-\eqref{wIana} (continuous curves) when $M \omega \ll 1$.
The cusps in the imaginary part of the frequency correspond to the threshold of the 
ergoregion instability, above which the QNMs turn from stable to unstable. \comment{Note 
that the instability is still present in the
extremal Kerr case ($a=M$) and continuously matches the instability of
superspinars ($a>M$)~\cite{Gimon:2007ur,Cardoso:2008kj,Pani:2010jz,Maggio:2017ivp}. The 
analytical approximation breaks down in the high-spin regime since $\omega_R M\sim 
m\Omega M\approx m/2$ is not small.}
}
    \label{fig:comparison}
\end{figure*}

%%%%%%%%%%%%%%%%%%%%%%%%%%%%%%%%%%%%%%%%%%%%%%%%%%%%
%
%
Figure~\ref{fig:comparison} shows the agreement between the QNMs
computed numerically and analytically through a matched asymptotic expansion.
As expected, the agreement is very good in the small-frequency regime, i.e.
when both $\log\epsilon$ and $\Omega M$ are small. In particular, the
agreement is excellent near the critical value of the spin $\chi_{\rm crit}$
(since $\omega=0$ at the critical value). When $a\approx 0$, the agreement
improves as $\epsilon\to0$ since in such limit $\omega\to0$. On the other hand,
when $a\approx M$, $M\omega\to mM\Omega\approx m/2$ as $\epsilon\to0$ and
therefore finite-frequency effects become important, especially for $m>1$.

Notice that both for axial and polar modes the imaginary part of the frequency
changes sign for a critical value of the spin, i.e., for $a > a_{\text{crit}}$
the ECO model becomes unstable against the ergoregion instability. \comment{The 
instability is still present in the extremal Kerr case ($a=M$) and in the superspinar case 
($a>M$). When $a= M$, the QNMs computed analytically are not reliable -- since they are 
not in the small-frequency regime -- thus we can rely only on the numerical results in 
order to estimate the instability in the extremal Kerr case. The instability in the 
superspinar case will be discussed in detail in a separated work. }

\begin{figure*}[th]
    \centering
    \includegraphics[width=0.49\textwidth]{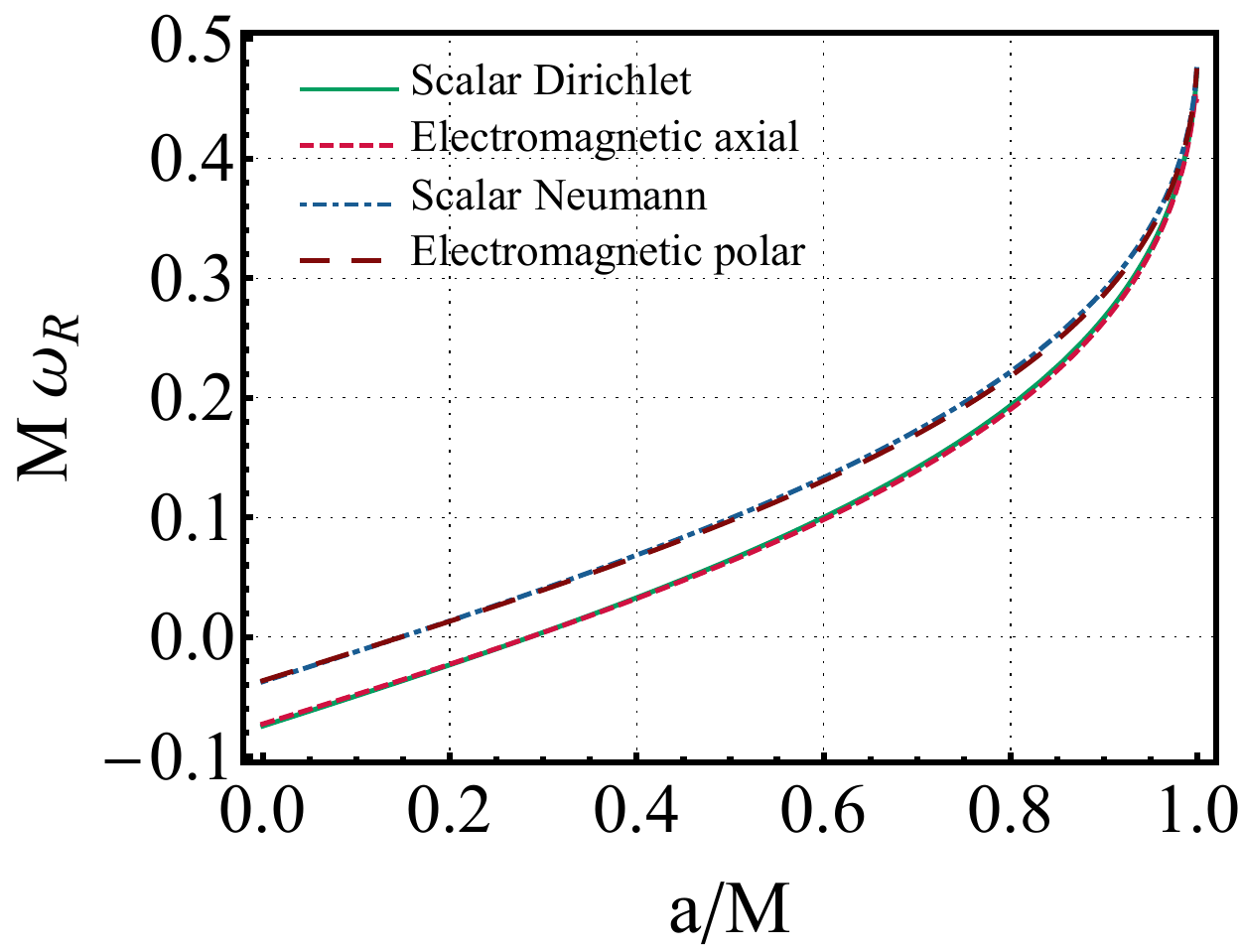}
    \includegraphics[width=0.49\textwidth]{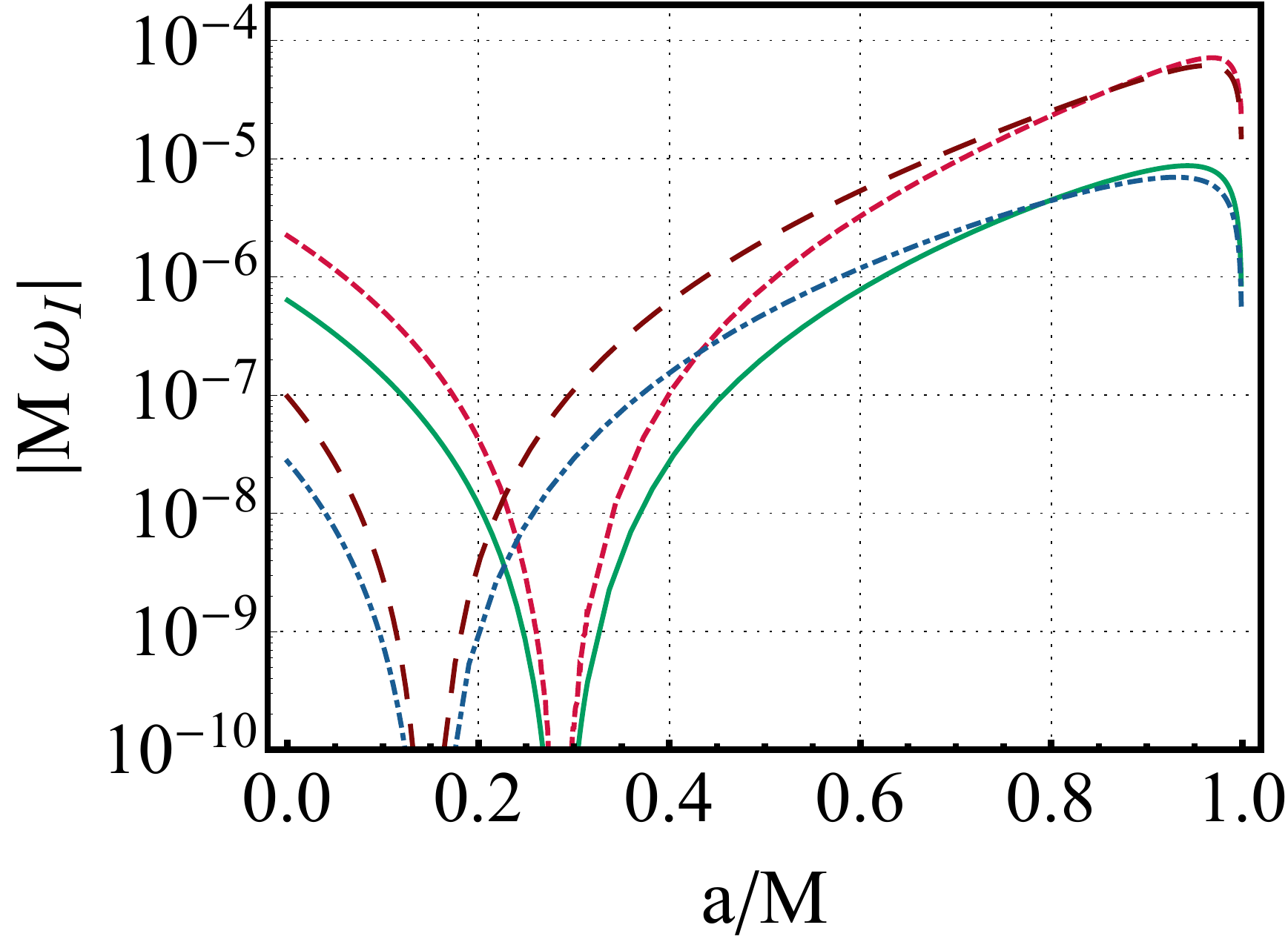}
    \caption{Real (left panel) and imaginary (right panel) part of the
fundamental scalar and electromagnetic QNMs of an ECO as a function of the spin,
where the surface of the ECO is at $r_0 = r_+ (1+\epsilon)$ with
$\epsilon=10^{-10}$. The cusps in the imaginary part of the frequency are the threshold of the ergoregion instability above which the QNMs becomes unstable.
The scalar QNMs with Dirichlet (Neumann) boundary condition
show a correspondence with the electromagnetic axial (polar) QNMs: the real part of the frequency tends to be the same for $\epsilon \ll 1$ and the critical value of instability is identical (within our numerical accuracy).
}
    \label{fig:scalem}
\end{figure*}

An interesting feature is that the threshold of instability is the same both for
scalar and electromagnetic perturbations within our numerical accuracy, as shown in the right panel of Fig.~\ref{fig:scalem}. In particular, scalar modes with Dirichlet (Neumann) boundary condition on Teukolsky's function $R_0$ turn unstable for the same critical value of the spin of electromagnetic axial (polar) modes.
In the next section we explain this finding analytically in terms of Darboux
transformations \cite{Glampedakis:2017rar} between perturbations of the Kerr metric with different $s$
index.

In the left panel of Fig.~\ref{fig:scalem}, we notice that the real part of scalar
QNMs with Dirichlet (Neumann) boundary condition and of
 electromagnetic axial (polar) QNMs tends to be the same in the BH limit, $\epsilon\to0$. 
This remark is confirmed by the analytical description of QNMs given in 
Eq.~\eqref{wRana}. According to the latter, the real part of the frequency is the same 
for $s=0$ and $s=-1$, and $\omega_R\sim m\Omega$ for $\epsilon \ll 1$ and any value of 
$s$.

Moreover, as shown in the right panel of Fig.~\ref{fig:scalem}, the imaginary part of the frequency displays a similar trend for scalar
Dirichlet (Neumann) QNMs and electromagnetic axial (polar) QNMs. The numerics is in 
agreement with Eq.~\eqref{wIana} according to which the imaginary part of the frequency in 
the electromagnetic case is the same as the scalar one multiplied by a factor 
$\beta_{1l}/\beta_{0l}$.

We also notice that the electromagnetic axial and polar modes tend to be the same in
the BH limit, as it happens for the scalar QNMs with Dirichlet
and Neumann boundary conditions in the same limit~\cite{Maggio:2017ivp}. This feature can be understood analytically by noticing that, as
$\epsilon\to0$, Eq.~\eqref{BCr0} reduces to
%%%
\begin{equation}
 \frac{dR_{-1}}{dr_*}=i\tilde\omega R_{-1}\,,
\end{equation}
%%%
for both axial and polar modes. Therefore, in the BH limit axial and polar
electromagnetic modes of the ECO are isospectral. Given that this is the
case for a BH~\cite{Chandra}, it is natural to conjecture that isospectrality
in the BH limit is a generic feature for any type of perturbation.
%

%%%%%%%%%%%%%%%%%%%%%%%%%%%%%%%%%%%%%%%%%%%%%%%%%%%%%%%%%%%%%%%%%%%%%%%%%%%%%%%%%%
\section{Static modes and Darboux transformations} \label{sec:zerofreq}
%%%%%%%%%%%%%%%%%%%%%%%%%%%%%%%%%%%%%%%%%%%%%%%%%%%%%%%%%%%%%%%%%%%%%%%%%%%%%%%%%%
%
The zero-frequency modes are associated with the onset of the ergoregion
instability~\cite{Brito:2015oca,Maggio:2017ivp}, for
reflecting boundary conditions. Here we seek a relationship between $\epsilon$,
the compactness parameter, and $a_{\text{crit}}$, the critical value of the
spin
at which the instability appears.

For $\omega = 0$, Teukolsky's equation (\ref{wave_eq}) reduces to the radial
equation
\beq
&& \Delta^{-s}\frac{d}{dr}\left(\Delta^{s+1}\frac{dR_s}{dr}\right) \nn \\
&& \quad +\left(\frac{a^2m^2+2is(r-M)am}{\Delta}-\lambda\right)R_s=0 \,,
\label{eq:static}
\eeq
where $\lambda=(l-s)(l+s+1)$.

%%%%%%%%%%%%%%%%%%%%%%%%%%%%%%%%%%%%%%%%%%%%%%%%
\subsection{Zero frequency modes: scalar field}
%%%%%%%%%%%%%%%%%%%%%%%%%%%%%%%%%%%%%%%%%%%%%%%%

For $s = 0$, Eq.~(\ref{eq:static}) has the general solution
\be
R_{0} = c_P P_l^{i \nu}(2 x + 1) + c_Q Q_l^{i \nu}(2 x + 1) \,,
\ee
where $\nu \equiv 2 a m / (r_+ - r_-)$ and $x \equiv (r - r_+)/(r_+ - r_-)$.
Here $P_l^{i\nu}(\cdot)$ and $Q_l^{i \nu}(\cdot)$ are associated Legendre
functions with the branch cut along the real axis from $-\infty$ to $1$. The
boundary condition of regularity as $r \rightarrow \infty$ imposes that $c_P =
0$. At the surface $r= r_0$, we impose
totally-reflecting (Dirichlet or Neumann) boundary conditions. That is,
\be
Q_l^{i\nu}(1 + 2 x_0) = 0 \quad \text{or} \quad
\left. \frac{d}{dx} Q_l^{i\nu}(1 + 2 x) \right|_{x=x_0} = 0 \,,
\label{eq:static-bc}
\ee
where $x_0 = \epsilon r_+ / (r_+ - r_-)$. By solving Eq.~(\ref{eq:static-bc})
numerically, we obtain the relationship between $\epsilon$ and the value
of $a$ for which a static mode occurs.

For ultracompact objects characterized by $\epsilon \ll 1$, it is
appropriate to use
\be
Q^{i\nu}_l \approx  \frac{e^{-\pi \nu}}{2\Gamma(i\nu)} \left[ x^{-i\nu/2} +
\frac{\Gamma(-i\nu)\Gamma(l+1+i\nu)}{\Gamma(i\nu)\Gamma(l+1-i\nu)}  x^{i \nu /
2} \right] \,,
\ee
leading to
\be
\ln \frac{\epsilon r_+}{r_+-r_-}  \approx - \frac{\pi (p + 1)}{\nu} +
\frac{i}{\nu} \ln
\frac{\Gamma(1-i\nu)\Gamma(l+1+i\nu)}{\Gamma(1+i\nu)\Gamma(l+1-i\nu)} \,,
\label{eq:static-approx}
\ee
where $p$ is a non-negative even (odd) integer for Neumann (Dirichlet)
modes. This makes it straightforward to find   the relationship between $\ln
\epsilon$ and $\nu = 2 a_{\text{crit}} m / (r_+ - r_-)$, and thus
$a_{\text{crit}}$, the critical value of $a$ at the threshold of the ergoregion
instability.

\begin{figure}[th]
    \centering
    \includegraphics[width=0.46\textwidth]{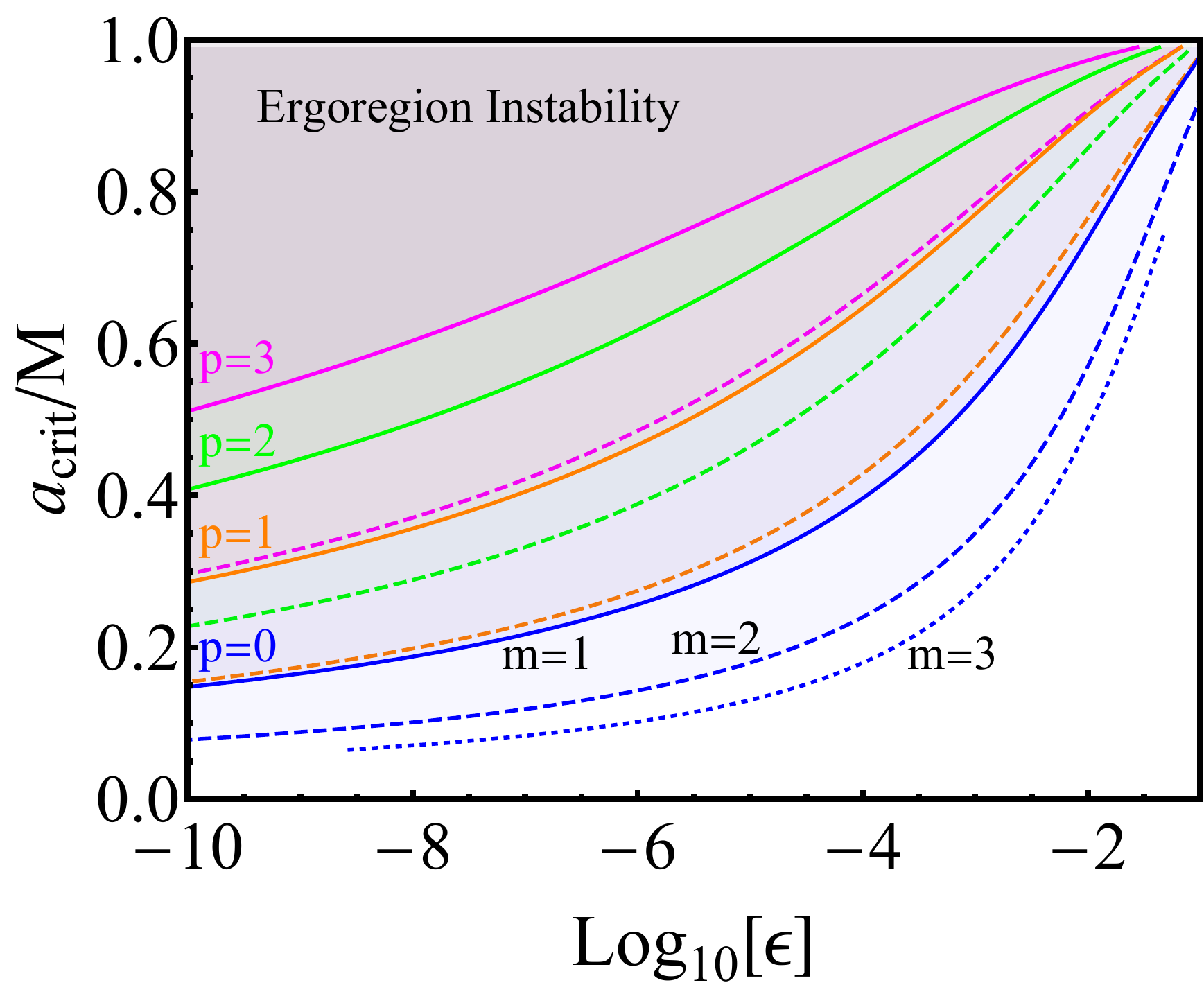}
    \caption{The ergoregion instability in the ECO spin-compactness parameter
space. The lines show the zero-frequency scalar QNMs in the $(\epsilon,
a_{\text{crit}})$ domain with Neumann ($p$ even) and Dirichlet ($p$ odd)
boundary conditions on the ECO surface at $r_0 = r_+ (1 + \epsilon)$. The
solid, dashed and dotted lines show the $m=1$, $2$ and $3$ modes with $l=m$,
respectively. The shaded regions indicate where the corresponding modes suffer
the ergoregion instability.}
    \label{fig:zero-modes}
\end{figure}

Figure \ref{fig:zero-modes} shows the zero-frequency modes in the $(\epsilon,
a_{\text{crit}})$ domain, for the Neumann ($p$ even) and Dirichlet ($p$ odd)
boundary conditions, for the $l=m$ modes with $m=1$ (solid), $m=2$ (dashed) and
$m=3$ (dotted). The plot shows that the ergoregion instability afflicts
co-rotating modes of the field. For each $m > 0$ there is a minimum value
$a_{\text{crit}}$ below which the mode is stable. Let us notice that $a_{\text{crit}}$
decreases as $m$ increases, and as $\epsilon$ decreases, so that even
slowly-rotating ECOs can suffer an ergoregion
instability~\cite{Maggio:2017ivp}, in principle. (For highly-spinning ECOs see Refs.~\cite{Hod:2017cga,Hod:2017wks}.)

In the limit $a \rightarrow 0$ and $\epsilon \rightarrow 0$, Eq.~(\ref{eq:static-approx}) reduces to
\beq
a_{\text{crit}} \approx \frac{\pi (p+1)}{m |\log \epsilon|} M \,,
\eeq
which is analogous to Eq.~\eqref{chicrit} derived analytically in the small-frequency regime.
So, for example, a totally reflecting barrier at the Planck scale outside the horizon of a 
$10M_{\odot}$ BH ($\epsilon = l_P / r_+ \sim 5 \times 10^{-40}$) will generate an 
ergoregion instability in the dipole mode ($m=1$) if $a \gtrsim 0.035$.

%%%%%%%%%%%%%%%%%%%%%%%%%%%%%%%%%%%%%
\subsection{Darboux transformations}
%%%%%%%%%%%%%%%%%%%%%%%%%%%%%%%%%%%%%

Now let us consider electromagnetic and gravitational perturbations.
It is straightforward to verify that, for $\omega=0$, the radial functions
$R_s$ are related to one another through the following transformations,
\begin{subequations}
\beq
R_{-1} &=& R_0+\frac{i\Delta}{am}\,R_0'\,,\\
R_{-2} &=& \frac{2a^2m^2-l(l+1)\Delta-2iam(r-M)}{am(2am-i(r_+-r_-))}R_0 \nn\\
&&+\frac{2\Delta(iam+r-M)}{am(2am-i(r_+-r_-))}R_0'\,, \label{darboux_grav}
\eeq
\end{subequations}
defined up to multiplication by a constant factor.
Though this relation is not unique, it seems to be the unique transformation
for which the fields are regular at large distances.
Equivalent transformations are
\begin{subequations}
\beq
R_0&=& - \frac{iam}{l(l+1)} \left[ R_{-1}' + \frac{iam}{\Delta} R_{-1} \right]
\,, \label{eq:R0darboux} \\
R_{-2}&=&\frac{a m-2i(r-M)}{2am-i(r_+-r_-)}R_{-1}+\frac{i\Delta}{2am-i(r_+-r_-)}R_{
-1}'\,. \nn \\
\eeq
\end{subequations}

%%%%%%%%%%%%%%%%%%%%%%%%%%%%%%%%%%%%%%%%%%%%%%%%%%%%%%%%%%%%%%%%%%%%%%
\subsection{Zero frequency modes: electromagnetic \& gravitational
perturbations}
%%%%%%%%%%%%%%%%%%%%%%%%%%%%%%%%%%%%%%%%%%%%%%%%%%%%%%%%%%%%%%%%%%%%%%

In the zero-frequency limit, the boundary condition (\ref{BCr0}) on the electromagnetic wavefunction reduces to
\be
R^\prime_{-1} + i \left( \frac{a m}{\Delta} - \frac{\varsigma_{\pm}  l
(l+1)}{am} \right) R_{-1} = 0 \,, \label{eq:bcstatic}
\ee
where $\varsigma_+ = 1$ for polar modes and $\varsigma_- = 0$ for axial modes. By comparison with
Eq.~(\ref{eq:R0darboux}), we see that axial modes are generated from a
scalar-field solution with a Dirichlet boundary condition $R_0(r_0) = 0$. For
the polar modes, we may take a derivative of (\ref{eq:R0darboux}) and
use the static Teukolsky equation (\ref{eq:static}) to establish that
\be
\Delta R_0' = \frac{a^2m^2}{l(l+1)} \left[R_{-1}^\prime + i
\left(\frac{am}{\Delta} - \frac{l(l+1)}{am} \right) R_{-1}\right] \,.
\ee
Thus, from Eq.~(\ref{eq:bcstatic}), the polar modes are generated from a
scalar-field solution with a Neumann boundary condition\footnote{In 
Ref.~\cite{Maggio:2017ivp} the Neumann boundary condition is
imposed on the radial wavefunction $Y_0$ defined as $Y_0=(r^2 + a^2)^{1/2} R_0$.
However, in the small-$\epsilon$ limit, the QNM spectrum is
analogous to the spectrum obtained by imposing a Neumann boundary condition on
$R_0$.} $R_0'(r_0) = 0$. 
It is natural to posit that this
extends to gravitational perturbations as well, for which case one has the following boundary conditions in the zero-frequency limit
\begin{subequations}\label{BCs2Darboux}
\beq
R^\prime_{-2}&=&-\left[\frac{(l-1)(l+2)}{2(iam+r-M)}+\frac{iam}{\Delta}\right]R_{-2}\,,\\
 R^\prime_{-2}&=&-\frac{iam(l-1)(l+2)}{2iam(iam+r-M)+l(l+1)\Delta}R_{-2}\non\\
&&-\frac{iam}{\Delta}R_{-2}\,,
\eeq
\end{subequations}
which are generated from a scalar-field solution with Dirichlet and Neumann boundary conditions, respectively.

By virtue of the Darboux transformations, it follows that
Eqs.~(\ref{eq:static-bc}) and (\ref{eq:static-approx}), and
Fig.~\ref{fig:zero-modes}, fully describe the zero-frequency modes of, not just
a scalar field, but also electromagnetic perturbations for an object with
perfectly reflecting boundary conditions. (Note however that the critical spin is slightly different in the gravitational case, as discussed in the next section, see Eq.~\eqref{chicrit_grav}.)

To the best of our knowledge these properties of static perturbations of the
Kerr metric have not been presented before. In particular, the above relations
show that the $\omega\to0$ limit of generic spin-$s$ perturbations is
universal.

%%%%%%%%%%%%%%%%%%%%%%%%%%%%%%%%%%%%%%%%%%%%%%%%%%%%%%%%%%%%%%
\section{ECO instabilities for gravitational perturbations} \label{sec:grav}
%%%%%%%%%%%%%%%%%%%%%%%%%%%%%%%%%%%%%%%%%%%%%%%%%%%%%%%%%%%%%%

In principle, gravitational perturbations of our Kerr-like ECO model can be
studied by solving Teukolsky's equation~\eqref{wave_eq} [or its alternative
version in Detweiler's form~\eqref{final}] with $s=\pm2$. However,
in this case the
issue of boundary conditions is much more subtle (see Ref.~\cite{Price:2017cjr}
for a related discussion). In the electromagnetic case,
the boundary conditions~\eqref{BCr0} are derived by assuming that the object is
made of a conducting material, so that the two boundary conditions in
Eq.~\eqref{BCr0} are ultimately related to the requirement that the electric and
magnetic field be orthogonal and parallel to the surface,
respectively~\cite{Brito:2015oca}.
An analog equation for the gravitational case is currently not
available.
Furthermore, in analogy with the electromagnetic case, the boundary
condition must depend on the properties of
the object's surface, which are also unknown and generically model-dependent.

Nevertheless, we now argue that the results of the previous section
can be easily extended to the gravitational case when $\epsilon\to0$, at least
if the object is perfectly reflecting.

%%%%%%%%%%%%%%%%%%%%%%%%%%%%%%%%%%%%%%%%%%%%%%%%%%%%%%%%%%%%%%%%%%%
\subsection{Analytical results} \label{sec:anagrav}
%%%%%%%%%%%%%%%%%%%%%%%%%%%%%%%%%%%%%%%%%%%%%%%%%%%%%%%%%%%%%%%%%%%
The previous analytical computation for the electromagnetic case can be
easily understood from a ``bounce-and-amplify''
argument~\cite{Brito:2015oca,Cardoso:2017njb}, i.e. in terms of quasi-standing
waves of a reflecting cavity \comment{between the surface of the ECO and the photon-sphere. Then the waves slowly leak out through tunneling in the photon-sphere barrier.} The frequency of (co-rotating) modes is set by the width
of the cavity, i.e.~$\omega_R-m\Omega\sim \pi/r_*^0$, whereas the instability is
controlled by the amplification factor $Z$ of the ergoregion at this
frequency~\cite{Brito:2015oca}, i.e. $\omega_I\sim Z \omega_R$.
\comment{A right-moving wave originating at the horizon has the asymptotics given by Eq.~\eqref{asymp_sol}. When it is backscattered by the photon-sphere, it acquires a factor $A_-$, where $|A_-| = |A_+|$ due to the Wronskian relationships \eqref{wronskian1}, \eqref{wronskian2}.
After this, the left-moving wave is further reflected at the surface of the ECO and is 
backscattered at the photon sphere again. At each following bounce at the photon sphere, 
it acquires a factor $A_-$. In turn, the factor $A_+$ describes the backscattering of a 
wave originating at infinity [Eq.~\eqref{asymp_sol}] and is related to the superradiant 
amplification factor of BHs through Eq.~\eqref{Z}. Thus, at each bounce the wave is 
amplified by a factor $Z$.
This argument applies to perfectly reflecting compact objects, since in this case the 
energy is conserved near the surface during subsequent bounces.}

A more quantitative way to look at this effect is to notice that the boundary
conditions~\eqref{BCr0} reduce to Dirichlet and Neumann boundary conditions for
the Detweiler function $X^-_s$. As we have shown, this is true for both scalar
($s=0$) and
electromagnetic ($s=\pm1$) perturbations. It is now natural to conjecture that
the same is true for gravitational perturbations, namely that Dirichlet and
Neumann conditions on $X^-_{\pm 2}$ imply perfect reflection at the
surface (i.e., no absorption by the interior). With this working assumption,
the analytical derivation of Sec.~\ref{subsec:analyticem} follows 
straightforwardly and the
final result in Eqs.~\eqref{wRana} and~\eqref{wIana} is valid also for
low-frequency gravitational perturbations $(s=\pm2$), the only difference due to the phase 
of $\omega_R$ computed in Appendix~\ref{app:matchings2}, namely 
%%%
\comment{ 
\begin{eqnarray}
\omega_R&\sim&
- \frac{\pi (q+1)}{2|r_*^0|}+m\Omega \,,
\label{wRana_grav} \\
 %%%
 \omega_I &\sim&
-\frac{\beta_{2l}}{|r_*^0|}\left(\frac{2 M r_+}{r_+-r_-}\right)\left[
\omega_R(r_+-r_-)\right]^{2l+1}(\omega_R-m\Omega)\,, \nn\\\label{wIana_grav}
\end{eqnarray}
}
%%%
where $q$ is a positive odd (even) integer for polar (axial) modes.
\comment{Note that the above result is a particular case of Eqs.~\eqref{wRana} and \eqref{wIana} 
for $s=-2$. In particular, in the gravitational case the real part of the frequency has a 
factor $\pi$ of difference in the phase compared to the scalar and electromagnetic cases.}

Equation~\eqref{wRana_grav} implies that the gravitational modes become 
unstable for a critical value of the spin which is different from the electromagnetic 
case, in particular \comment{(for $s=-2$)}
\comment{
\begin{equation}
 \chi>\chi_{\rm crit} \approx\frac{\pi (q+1)}{m |\log\epsilon|}\,.
\label{chicrit_grav}
\end{equation}
}
For example, for $\epsilon = 10^{-10}$ \comment{and $l=m=2$} $\chi_{\rm crit, em} \approx 0.14$ whereas \comment{
$\chi_{\rm crit, grav} \approx 0.20$} for axial modes, and $\chi_{\rm crit, em} \approx 
0.07$ whereas \comment{$\chi_{\rm crit, grav} \approx 0.14$} for polar modes.

Furthermore, in the $\epsilon\to0$ limit ($\omega_R\to m\Omega$),
the instability time scale is proportional to \comment{the inverse of the term
$\beta_{sl}[m\Omega(r_+-r_-)]^{2l+1}[q+s(s+1)/2]$ in $\omega_I$.}
Since $\beta_{11}=4\beta_{01}$, the instability time scale for electromagnetic
perturbations is simply four times shorter than for scalar perturbations.
On the other hand, for gravitational perturbations the dominant mode has
$|s|=l=m=2$, which gives $\beta_{22}=\beta_{11}/25$. \comment{Taking into account also
the $[m\Omega(r_+-r_-)]^{2l+1}[q+s(s+1)/2]$ term, in the $\epsilon\to0$ limit we obtain
%%%
\begin{equation}
 \tau_{222}=\frac{25(1+\sqrt{1-\chi^2})^2 q}{32\chi^2(1-\chi^2) (1+q)} \tau_{111}\,,
\end{equation}
}
%%%
where $\tau_{slm}=1/\omega_I$ for a given $(s,l,m)$. Note that
$\tau_{222}>\tau_{111}$ for any spin, showing that the dominant ($l=2$)
gravitational instability is actually \emph{weaker} than the dominant ($l=1$)
electromagnetic
instability in the low-frequency limit. This is consistent with the fact that
the amplification factor for $s=l=2$ waves is smaller than that
of $s=l=1$ waves at low frequency~\cite{Brito:2015oca} (see also
Fig.~\ref{fig:Z} below).

Finally, our conjecture is also supported by the zero-frequency limit
discussed in Sec.~\ref{sec:zerofreq}, where we showed that the behavior for
scalar, electromagnetic and gravitational perturbations is universal in the
zero-frequency limit. Indeed, the boundary conditions~\eqref{BCsXgrav} 
respectively reduce to Eqs.~\eqref{BCs2Darboux} for $\omega\to0$ (which also requires 
$a\to a_{\rm crit}$) and $\epsilon\to0$.

%%%%%%%%%%%%%%%%%%%%%%%%%%%%%%%%%%%%%%%%%%%%%%%%%%%%%%%%%%%%%%%%%%%
\subsection{Numerical results}
%%%%%%%%%%%%%%%%%%%%%%%%%%%%%%%%%%%%%%%%%%%%%%%%%%%%%%%%%%%%%%%%%%%
\label{subsec:numericsgrav}

%%%%%%%
\begin{figure*}[th]
    \centering
    \includegraphics[width=0.45\textwidth]{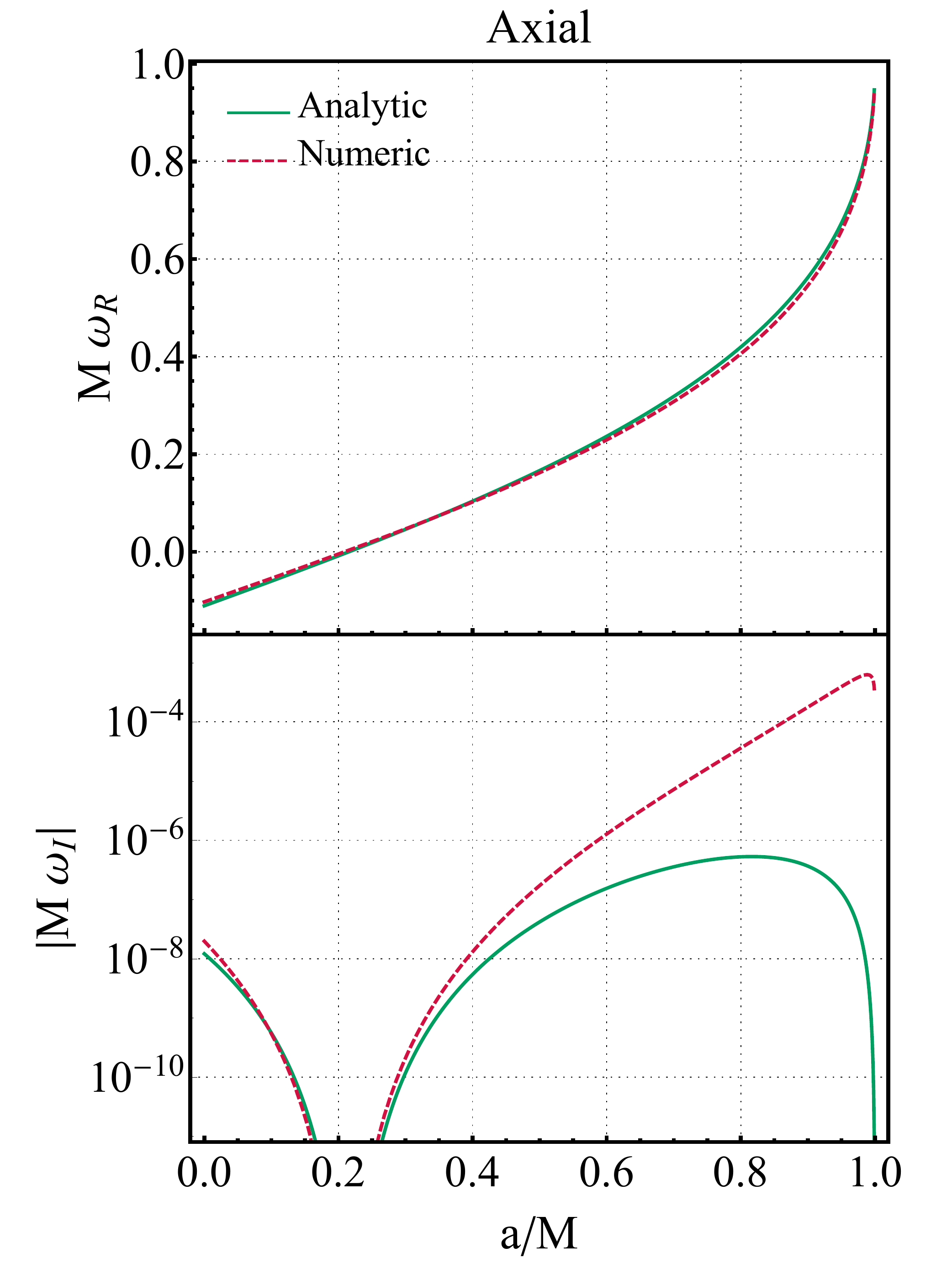}
    \includegraphics[width=0.45\textwidth]{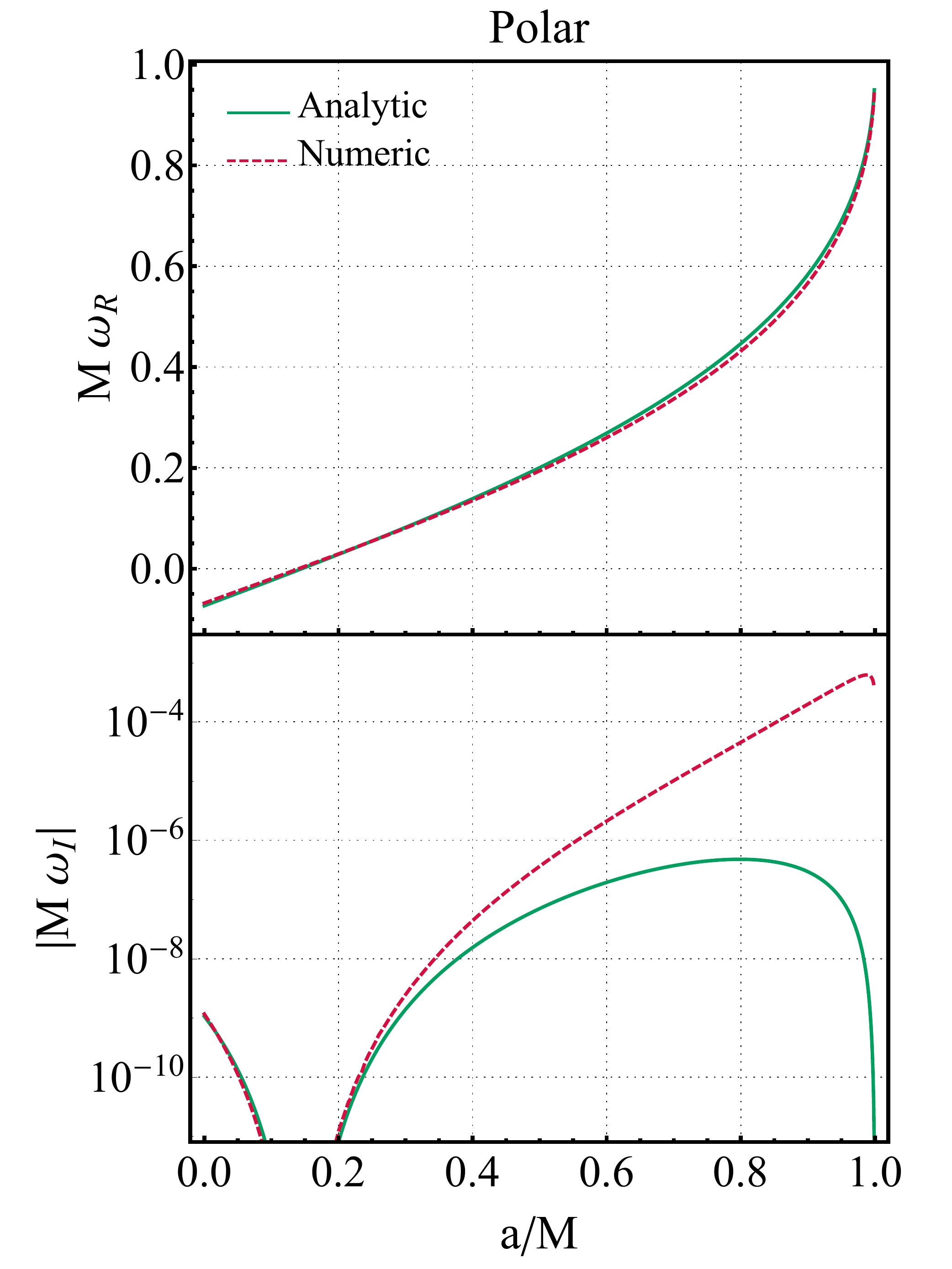}
    \caption{Real (top panels) and imaginary (bottom panels) part of the fundamental
gravitational QNM ($l=m=2$, $n=0$) of an ECO as a function of the spin. The left (right) panels refer to axial (polar) modes corresponding to Dirichlet (Neumann) boundary condition on Detweiler's function $X_{-2}$, and the surface of the ECO is
at
$r_0 =
r_+ (1+\epsilon)$ with $\epsilon = 10^{-10}$. The QNMs computed numerically
(dashed curves) are in agreement with the QNMs computed analytically through
Eqs.~\eqref{wRana_grav}-\eqref{wIana_grav} (continuous curves) when $M \omega \ll 1$. \comment{In the extremal Kerr case ($a=M$) the analytical approximation is not valid since $M \omega = \mathcal{O}(1)$.}
}
    \label{fig:comparison_grav}
\end{figure*}
%%%%%%
We solve Teukolsky's equation~\eqref{wave_eq}
for gravitational perturbations numerically by assuming Dirichlet and Neumann boundary conditions
for the Detweiler's function $X_{-2}$ (see Eqs.~\eqref{BCsXgrav}). A representative example is shown in \comment{Fig.~\ref{fig:comparison_grav}}, where we present the fundamental ($n=0$) $l=m=2$ gravitational modes of an ECO as a function of the spin.
The qualitative behavior of the modes is very similar to the electromagnetic case 
(compare Fig.~\ref{fig:comparison_grav} with Fig.~\ref{fig:comparison}) and the agreement 
with the analytical result derived in the previous section is very good in the regime 
where $M\omega\ll1$, i.e. near the threshold of the instability.

Note that in this case the frequency is overall larger, since $\omega_RM\to 2\Omega M\approx 1$ for $m=2$ and $\chi\to1$. In this limit, the quantity $a\omega\approx 1$ and the angular eigenvalues need to be computed numerically.

%%%%%%%%%%%%%%%%%%%%%%%%%%%%%%%%%%%%%%%%%%%%%%%%%%%%%%%%%%%%%%%%%%%
\section{Discussion: the role of absorption and astrophysical
implications} \label{sec:absorption}
%%%%%%%%%%%%%%%%%%%%%%%%%%%%%%%%%%%%%%%%%%%%%%%%%%%%%%%%%%%%%%%%%%%

Owing to the logarithmic dependence in Eq.~\eqref{wIana}, the
ergoregion-instability time scale for perfectly-reflecting ECOs is always
very short, even for Planck-inspired objects with $\epsilon\sim{\cal
O}(10^{-40})$~\cite{Maggio:2017ivp}. The most natural development of the
instability is to remove angular momentum until the superradiant condition is
saturated~\cite{Cardoso:2014sna}. Thus, spin measurements of dark compact
objects indirectly rule out perfectly-reflecting ECOs. Furthermore, the absence
of any detectable gravitational-wave stochastic background in LIGO
O1~\cite{TheLIGOScientific:2016wyq,TheLIGOScientific:2016dpb} sets the most
stringent constraints to date on these models~\cite{Barausse:2018vdb}. \comment{(It is worth mentioning that the stochastic gravitational-wave background specifically from the spin-down of remnants of binary mergers is not inconsistent with gravitational wave observations, as discussed in
Ref.~\cite{Fan:2017cfw}.)}

However, the instability can be totally quenched by (partial) absorption by the object interior~\cite{Maggio:2017ivp}.
In the case of scalar perturbations, this requires
reflectivity $\sim0.4\%$ smaller than unity~\cite{Maggio:2017ivp}. This number corresponds to
the maximum superradiant amplification factor for scalar perturbations of a
Kerr BH~\cite{Press:1972zz,Brito:2015oca} and it is indeed consistent with the
``bounce-and-amplify'' argument presented in the previous section. Namely, from
Eq.~\eqref{asymp_sol} a right-moving monochromatic wave is backscattered by the
ECO potential and acquires a factor $A_-=A_+$. The reflected wave travels to the
left and is further reflected at the surface. Let us assume that the reflection
coefficient (i.e., the ratio between the outgoing energy flux to
the ingoing energy flux) at the object's surface is $|{\cal
R}(\omega)|^2$. Then, the
left-moving wave $A_+ e^{-i\tilde\omega r_*}$ is reflected at the surface as
$A_+{\cal R} e^{i\tilde\omega r_*}$ (see~\cite{Testa:2018bzd} for a model based
on geometrical optics and Ref.~\cite{Mark:2017dnq} for a generic
discussion). The process continues indefinitely and the wave
acquires a factor $A_+{\cal R}$ for each bounce. Therefore, the condition for
the energy in the cavity to grow indefinitely in time is $|A_+{\cal R}|^2>1$ or
%%%
\begin{equation}
 |{\cal R}|^2>\frac{1}{1+Z}\,,\label{condition}
\end{equation}
%%%
where both the amplification factor $Z=|A_+|^2-1$ and the ECO reflection coefficient $|{\cal R}|^2$ are evaluated at the dominant frequency $\omega=\omega_R$. In the low-frequency regime, $Z$ is approximately given by Eq.~\eqref{amplfactor}, but it can be computed numerically for any frequency and spin~\cite{Teukolsky:1974yv,Brito:2015oca}.
Since $|{\cal R}|^2\leq1$, Eq.~\eqref{condition} implies that a necessary condition for the instability is $Z>0$, i.e.~the relevant frequency needs to be in the superradiant regime to trigger the instability. Furthermore, if the object is perfectly reflecting ($|{\cal R}|^2=1$), the instability is quenched only when $Z<0$, i.e., only in the absence of superradiance.
Likewise, if the object is almost a BH (${\cal R}\approx0$) the instability is absent for any finite amplification factor $Z$.

Equation~\eqref{condition} also implies that, in order to quench the instability
completely, it is sufficient that $|{\cal R}|^2>1/(1+Z_{\rm max})\approx
1-Z_{\rm max}$, where $Z_{\rm max}$ is the maximum amplification
coefficient\footnote{A less stringent condition is  $|{\cal
R(\omega_R)}|^2>1/(1+Z(\omega_R))$, where $\omega_R$ is the dominant QNM
frequency.}, and the last approximation is valid when $Z_{\rm max}\ll1$, as it
is typically the case~\cite{Teukolsky:1974yv,Brito:2015oca}.

%%%
\begin{figure}[th]
    \centering
    \includegraphics[width=0.55\textwidth]{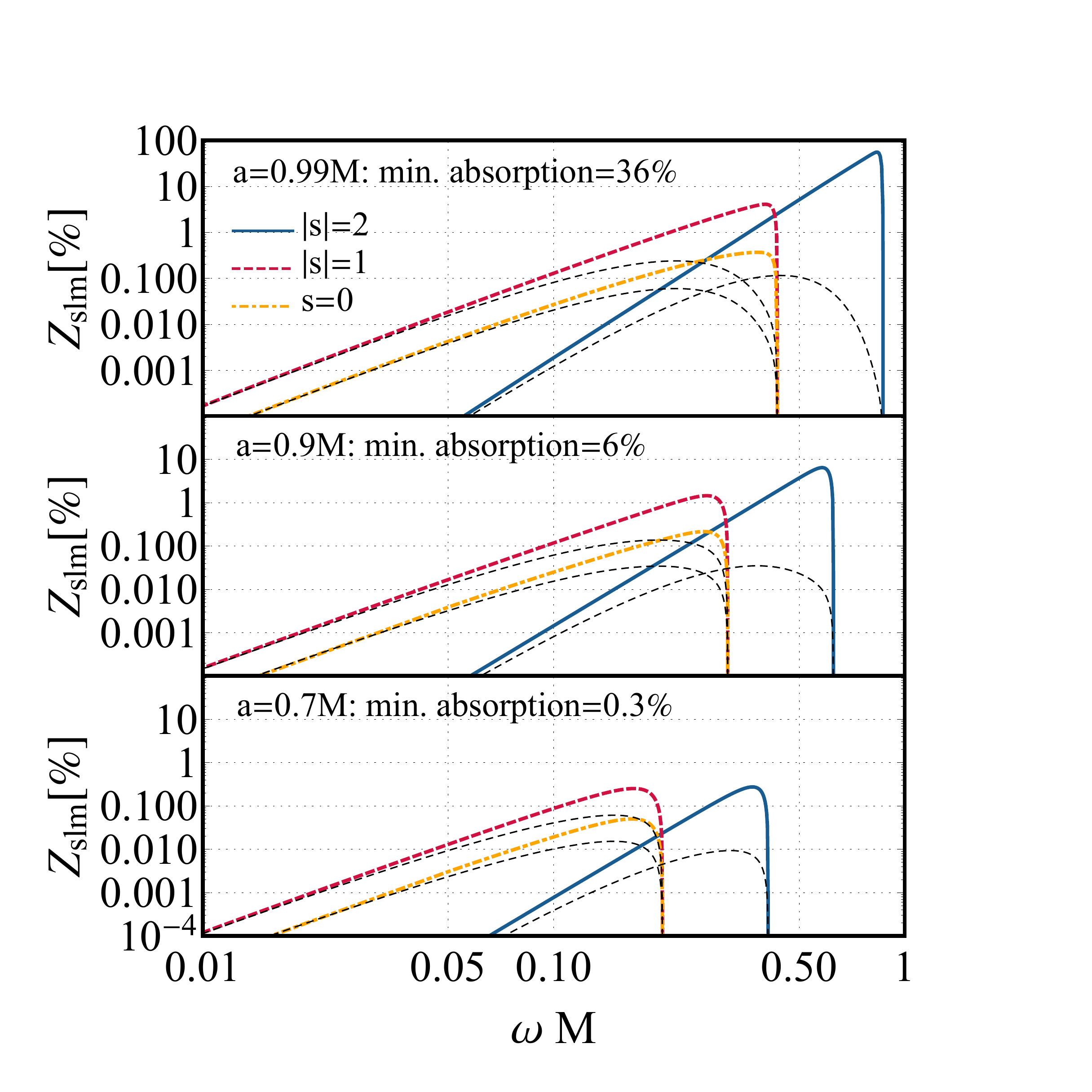}
    \caption{Superradiant amplification factor for a
Kerr BH as a function of the frequency $\omega$ of the incident wave for
different values of the BH spin and for different types of perturbations (we
set $l=m=1$ for scalar and electromagnetic perturbations, and $l=m=2$ for
gravitational perturbations). The analytical approximation~\eqref{amplfactor}
valid at low-frequency (black dashed lines) is compared to the exact numerical
result~\cite{Teukolsky:1974yv,Brito:2015oca}. In each panel we report the
minimum absorption coefficient at the ECO surface, $1-|{\cal R}|^2$, necessary to quench the instability, as obtained by saturating Eq.~\eqref{condition}.}
    \label{fig:Z}
\end{figure}
%%%%

The above discussion is consistent with the analysis of Ref.~\cite{Maggio:2017ivp} for scalar perturbations, but it is actually valid for any kind of perturbations of our ECO model in the BH limit.
In general, Eq.~\eqref{condition} implies that the larger the BH amplification factor the larger is the minimum absorption rate necessary to quench the instability.
The superradiant amplification factor (and hence the minimum absorption rate required to quench the instability) depends
significantly on the spin. This is shown in Fig.~\ref{fig:Z}, where we present $Z(\omega)$ for
different values of the spin of a Kerr BH and for different types of perturbations~\cite{Teukolsky:1974yv,Brito:2015oca}.
As predicted by the analytical result, electromagnetic perturbations have the largest amplification factor at low frequency. On the other hand, gravitational perturbations can be amplified much more than electromagnetic or scalar perturbations at high frequency which, by the superradiant condition $\omega(\omega-m\Omega)$ also require highly-spinning objects.
The minimum absorption coefficient, $1-|{\cal R}|^2$, required to quench the instability depends strongly on the spin. For an ECO spinning at $\chi\lesssim 0.9$ ($\chi\lesssim 0.7$), an absorption coefficient
of at least $6\%$ ($0.3\%$) is sufficient to quench the instability for any type of perturbation.
On the other hand, since the maximum superradiance amplification factor is
$\approx138\%$ (for $l=m=2$ gravitational perturbations of almost
extremal BHs~\cite{Teukolsky:1974yv,Brito:2015oca}), Eq.~\eqref{condition}
predicts that an aborption coefficient of at least $\approx60\%$ would quench the
instability in \emph{any} case.

In other words, if a specific ECO can be parametrized by an absorption
coefficient of (say) $\approx 1\%$ and it is  rapidly spinning when produced
(say, $\chi\approx1$), it would lose angular momentum over a short time scale
given by the ergoregion instability, until it reaches a critical value of the
spin corresponding to the saturation of Eq.~\eqref{condition}. From
Fig.~\ref{fig:Z}, in this example the saturation would roughly corresponds
to a final spin $\chi\approx0.8$.

%%%%%%%%%%%%%%%%%%%%%%%%%%%
\section{Conclusion}
%%%%%%%%%%%%%%%%%%%%%%%%%%%
We have computed the ergoregion instability for electromagnetic and gravitational perturbations
of a model of Kerr-like ECO, both numerically (for any compactness and spin)
and analytically (in the low-frequency regime, which is valid for small spin
and in the BH limit).
These cases are qualitatively similar to the
scalar case studied in Ref.~\cite{Maggio:2017ivp} and allow us to draw a
general picture of the ergoregion instability for ECOs. In particular, we
showed that: (i)~our analytical result can also be extended to the gravitational
perturbations of a perfectly-reflecting ECOs in the BH limit; (ii)~the
instability can be understood in terms of waves trapped within the photon-sphere
barrier and amplified by superradiant scattering~\cite{Brito:2015oca}.
Therefore, for any kind of perturbations the instability is completely quenched
if the absorption rate at the ECO surface is at least equal to the maximum
superradiance amplification for a given spin-$s$ perturbation of a Kerr BH with
same mass and spin; (iii)~the numerical results for both electromagnetic 
and gravitational perturbations agree well with the analytical ones in the small-frequency approximation.

As a by-product of our analysis, we have also found a set of Darboux
transformations that relate the waveforms of $s=0,\pm1,\pm2$ perturbations of
the Kerr metric in the static limit. It would be interesting to check whether
similar transformations exist also between bosonic and fermionic ($s=\pm1/2$)
perturbations. For the latter there is no superradiance~\cite{Brito:2015oca}
and therefore even highly-spinning, perfectly-reflecting ECOs should be stable
against $s=\pm1/2$ perturbations.

An interesting extension concerns superspinars, i.e. string-inspired,
regularized Kerr geometries spinning above the Kerr bound~\cite{Gimon:2007ur}.
The ergoregion instability of these models has been studied for scalar
perturbations~\cite{Cardoso:2008kj,Maggio:2017ivp} and for gravitational
ones~~\cite{Cardoso:2008kj,Pani:2010jz}. However, in the latter case the 
boundary conditions adopted do not correspond to Dirichlet or Neumann
conditions on the Detweiler function which, as we argued in this work, are the
appropriate ones for gravitational perturbations. In light of our results,
it would be interesting to extend the stability analysis of superspinars, both
for perfectly-reflecting and for partially-absorbing models.

%%%%%%%%%%%%%%%%%%%%%%%%%%%%%%%%%%%%%%%%%%%%%%%%%%%%%%%%%%%%%%%%%%%%%%
\begin{acknowledgments}
%%%%%%%%%%%%%%%%%%%%%%%%%%%%%%%%%%%%%%%%%%%%%%%%%%%%%%%%%%%%%%%%%%%%%%%
\comment{We thank Swetha Bhagwat for interesting discussions.}
The authors acknowledge networking support by the COST Action CA16104.
V.C. acknowledges financial support provided under the European Union's H2020 ERC Consolidator Grant ``Matter and strong-field gravity: New frontiers in Einstein's theory'' grant agreement no. MaGRaTh--646597.
P.P. acknowledges the financial support provided under the
European Union's H2020 ERC, Starting Grant agreement no.~DarkGRA--757480, and
and the kind hospitality of the Universitat de les Illes Balears, where this
work has been finalized.
S.D. acknowledges financial support from the European Union's Horizon 2020
research and innovation programme under the H2020-MSCA-RISE-2017 Grant
No.~FunFiCO-777740, and from the Science and Technology Facilities Council
(STFC) under Grant No.~ST/P000800/1.
\comment{E.M. and P.P.} acknowledge support from the Amaldi Research Center funded by the
MIUR program "Dipartimento di Eccellenza" (CUP: B81I18001170001). 
\end{acknowledgments}
%%%%%%%%%%%%%%%%%%%%%%%%%%%%%%%%%%%%%%%%%%%%%%%%%%%%%%%%%%%%%%%%%%%%%%%%%%%%%%
%
\appendix

%%%%%%%%%%%%%%%%%%%%%%%%%%%%%%%%%%%%%%%%%%%%%%%%%%%%%%%%%%%%%%%%%%%%%%%%%%%%%%%%%%%%%%%%%%%%%%%%%%%%
\section{Analytical asymptotic matching for spin-s perturbations}\label{app:matching}
%%%%%%%%%%%%%%%%%%%%%%%%%%%%%%%%%%%%%%%%%%%%%%%%%%%%%%%%%%%%%%%%%%%%%%%%%%%%%%%%%%%%%%%%%%%%%%%%%%%%
%
In this appendix we derive the electromagnetic and gravitational QNMs of an ECO 
analytically in the small-frequency regime through a matched asymptotic expansion.

In the region near the surface of the ECO, the radial wave equation~\eqref{wave_eq} reduces to \cite{Starobinskij2}
\begin{eqnarray}
&& \left[x(x+1)\right]^{1-s} \partial_x \left\{ \left[x(x+1)\right]^{s+1} \partial_x R_s \right\} \nn \\
&& \quad +\left[ Q^2 + i Q s (1+2x) - \lambda x (x+1)\right] R_s = 0 \,, \label{wave_eq_near_hor}
\end{eqnarray}
where $x=(r-r_+)/(r_+ - r_-)$, $Q=(r_+^2 + a^2)(m \Omega - \omega)/(r_+ - r_-)$, and 
$\lambda = (l-s)(l+s+1)$. The Eq.~\eqref{wave_eq_near_hor} is valid when $M \omega \ll 1$ 
and it is derived by neglecting the terms proportional to $\omega$ in Eq.~\eqref{wave_eq} 
except for the ones which enter into $Q$. The general solution of 
Eq.~\eqref{wave_eq_near_hor} is a linear combination of hypergeometric functions 
\beq
R_s &=& (1+x)^{iQ} \big[{C_1} x^{-iQ} \nn \\
&&_{2}F_1(-l+s,l+1+s;1-\bar{Q}+s;-x) + {C_2} x^{iQ-s} \nn \\
&&_{2}F_1(-l+\bar{Q},l+1+\bar{Q};1+\bar{Q}-s;-x)\big] \,, \label{sol_nearhor}
\eeq
where $\bar{Q}=2iQ$. The large-$r$ behavior of the solution is
\beq
R_s &\sim& \left(\frac{r}{r_+ - r_-}\right)^{l-s} \Gamma(2l+1) \Bigg[\frac{{C_1} \ 
\Gamma(1-\bar{Q}+s)}{\Gamma(l+1-\bar{Q}) \Gamma(l+1+s)} \nn \\
&+& \frac{{C_2} \ \Gamma(1+\bar{Q}-s)}{\Gamma(l+1+\bar{Q}) \Gamma(l+1-s)} \Bigg] + 
\left(\frac{r}{r_+ - r_-}\right)^{-l-1-s} \nn \\
& & \frac{(-1)^{l+1+s}}{2 \Gamma(2l+2)} \Bigg[\frac{{C_1} \ \Gamma(l+1-s) 
\Gamma(1-\bar{Q}+s)}{\Gamma(-l-\bar{Q})} \nn \\
&+& \frac{{C_2} \ \Gamma(l+1+s) \Gamma(1+\bar{Q}-s)}{\Gamma(-l+\bar{Q})} \Bigg] \,, 
\label{Rhorizon}
\eeq
where the ratio of the coefficients ${C_1}/{C_2}$ is fixed by the boundary condition at 
the surface of the ECO.

At infinity, the radial wave equation~\eqref{wave_eq} reduces to~\cite{Cardoso:2008kj}
\be
r \partial_{r}^2 f_{s} + 2 (l+1-i \omega r) \partial_{r} f_{s} - 2 i (l+1-s) \omega f_{s} 
= 0 \,, \label{wave_eq_inf}
\ee
where $f_{s} = e^{i \omega r} r^{-l+s} R_{s}$. The general solution of 
Eq.~\eqref{wave_eq_inf} is a linear combination of a confluent hypergeometric function 
and a Laguerre polynomial
\beq
R_{s} &=& e^{-i \omega r} r^{l-s} \big[ C_3 \ U(l+1-s,2l+2,2i \omega r) \nn \\
&+& C_4 \ L_{-l-1+s}^{2l+1}(2 i \omega r) \big] \,,
\eeq
where $C_4 = (-1)^{l-s} \ C_3 \ \Gamma(-l+s)$ by imposing only outgoing waves at 
infinity.
The small-$r$ behavior of the solution is
\beq
R_{s} &\sim& C_3 \ r^{l-s} \frac{(-1)^{l-s}}{2}  \frac{\Gamma(l+1+s)}{\Gamma(2l+2)} 
\nn \\
&+&  C_3 \ r^{-l-1-s} (2 i \omega)^{-(2l+1)} \frac{\Gamma(2l+1)}{\Gamma(l+1-s)} \,. 
\label{Rinfinity}
\eeq

The matching of Eqs.~\eqref{Rhorizon} and \eqref{Rinfinity} in the intermediate region 
yields
\be
\frac{{C_1}}{{C_2}} = -\frac{\Gamma(l+1+s)}{\Gamma(l+1-s)} \left[\frac{R_+ + i (-1)^l 
(\omega (r_+ - r_-))^{2l+1} L S_+}{R_- + i (-1)^l (\omega (r_+ - r_-))^{2l+1} L S_-} 
\right] \label{eq:ab1}
\ee
where
\beq
R_\pm &\equiv& \frac{\Gamma(1 \pm \bar{Q} \mp s)}{\Gamma(l + 1 \pm \bar{Q})}\,, \quad
S_\pm \equiv \frac{\Gamma(1 \pm \bar{Q} \mp s)}{\Gamma(-l \pm \bar{Q})}\,, \nn \\
L &\equiv& \frac{1}{2} \left[\frac{2^l \, \Gamma(l+1+s)\Gamma(l+1-s)}{\Gamma(2l+1) 
\Gamma(2l+2)} \right]^{2} \,.
\eeq

\subsection{Electromagnetic case}\label{app:matchings1}
For $s=-1$, the ratio ${C_1} / {C_2}$ is derived by imposing the boundary conditions 
\eqref{BCr0} in the near-horizon expansion of the solution in the near-horizon region. At 
the surface, we obtain 
\be
\frac{{C_1}}{{C_2}} = \mp B^{-1} \bar{Q} x_0^{\bar{Q}} \,. \label{eq:ab2}
\ee
where $x_0 = x(r_0)$, and the minus and plus signs refer to polar and axial 
perturbations, respectively.

By equating Eq.~(\ref{eq:ab1}) with Eq.~(\ref{eq:ab2}), we obtain an algebraic equation 
for the complex frequency $\omega$.
An approximate solution of $\omega$ can be found in the regime
$a\ll M$ and $\epsilon \ll 1$, i.e. $\bar{Q} \ll 1$. In this case, Eq.~\eqref{eq:ab1} 
reduces to ${C_1} / {C_2}= \bar{Q} / [l (l+1)]$, whereas $B \approx l (l+1)$ in 
Eq.~\eqref{eq:ab2}. It follows
\be
x_0^{-2iQ} = \mp 1 \,. \label{x0}
\ee
By using the tortoise coordinate $r_*^0 = r_* (r_0)$, where $ \log (x_0) \sim r_*^0 (r_+ 
- r_-)/(r_+^2 + a^2)$, Eq.~\eqref{x0} yields
\be
e^{-2 i Q r_*^0 (r_+ - r_-)/(r_+^2 + a^2)} = \mp 1 \,, \label{eqx0}
\ee
which is analogous to Eq.~(A18) in Ref.~\cite{Cardoso:2008kj} for the scalar-field case. 
The solution of Eq.~\eqref{eqx0} is
\be
\omega = - \frac{\pi q}{2 |r_*^0|} + m \Omega \,, \label{omega_em}
\ee
where $q$ is a positive odd (even) integer for polar (axial) modes. 
Equation~\eqref{omega_em} is also valid for scalar perturbations where $q$ is a 
positive odd (even) integer for the modes with \comment{Neumann (Dirichlet)} boundary condition on 
the Teukolsky's function.

\subsection{Gravitational case} \label{app:matchings2}
For $s=-2$, the ratio ${C_1} / {C_2}$ is derived by imposing the boundary conditions 
\eqref{BCsXgrav} in the near-horizon expansion of the solution in the near-horizon 
region. At the surface, \comment{when $\bar{Q} \ll 1$,} we obtain
\be
\comment{
\frac{{C_1}}{{C_2}} = \mp \frac{2}{(l+2)(l+1)l(l-1)} \bar{Q} x_0^{\bar{Q}} \,, 
\label{eq:ab2grav}
}
\ee
where the minus and plus signs refer to polar and axial perturbations, respectively.
When $\bar{Q} \ll 1$, Eq.~\eqref{eq:ab1} reduces to ${C_1} / {C_2}= -2\bar{Q} / 
[(l+2)(l+1)l (l-1)]$. By equating the latter equation with Eq.~(\ref{eq:ab2grav}), we get
\comment{
\be
x_0^{-2iQ} = \pm 1 \,, \label{x0grav}
\ee
whose solution is
\be
\omega = - \frac{\pi (q+1)}{2 |r_*^0|} + m \Omega \,, \label{omega_grav}
\ee
where $q$ is a positive odd (even) integer for polar (axial) modes. We conclude that in the gravitational case the frequency has an additional phase $\pi$ with respect to the scalar and electromagnetic case.}

%%%%%%%%%%%%%%%%%%%%%%%%%%%%%%%%%%%%%%%%%%%%%%
\section{Transformations of Teukolsky equation to a real potential}\label{app:transformation}
%%%%%%%%%%%%%%%%%%%%%%%%%%%%%%%%%%%%%%%%%%%%%%
In this appendix we revisit and extend the computation by Detweiler~\cite{1977RSPSA.352..381D} and derive the transformation of the Teukolsky's function to bring Eq.~\eqref{wave_eq} in a form like Eq.~\eqref{final} with a real potential.
In doing so, we correct some mistakes of Ref.~\cite{1977RSPSA.352..381D}. On the other hand, we note that the electromagnetic case presented in Ref.~\cite{1977RSPSA.352..381D} is correct and we refer the reader to the original work for the explicit transformation\footnote{To the best of our knowledge, Ref.~\cite{PhysRevD.41.374} reported a misprint in the effective potential~(B23) of Ref.~\cite{1977RSPSA.352..381D}, but did not provide the corresponding expressions for the functions $\alpha$ and $\beta$. In this Appendix, we correct several mistakes in Eqs.~(B3)-(B14) in Ref.~\cite{1977RSPSA.352..381D} and provide the explicit expressions of $\alpha$ and $\beta$\comment{, extending the calculations in Ref.~\cite{10.2307/79029} to gravitational perturbations.}}.

The Starobinskii identity for gravitational perturbations reads~\cite{Teukolsky:1974yv}
\be
\frac{1}{4} R_{2} = \mathscr{D} \mathscr{D} \mathscr{D} \mathscr{D}  R_{-2} \,, \label{staridentity}
\ee
where \comment{$\mathscr{D} = \partial_r - i K/\Delta$}. According to Eq.~\eqref{staridentity}, we can write
\be
R_{2} = \mathfrak{a} R_{-2} + \frac{\mathfrak{b}}{\Delta} \frac{dR_{-2}}{dr} \,,
\ee
where
\beq
\mathfrak{a} &=& (a_1 + i a_2) \,, \\
\mathfrak{b} &=& i b_2 \,,
\eeq
and
\beq
\nn a_1 &=& 4 \left[ \frac{8 K^4}{\Delta^4} + \frac{8 K^2}{\Delta^3} \left( \frac{M^2-a^2}{\Delta} - \lambda\right) \right.\\
\nn &-& \left. \frac{4 \omega K}{\Delta^3} (3r^2+2Mr-5a^2)+\frac{12r^2 \omega^2 + \lambda (\lambda+2)}{\Delta^2} \right]\,, \\ \\
\nn a_2 &=& 4 \bigg\{ - \frac{24 \omega r K^2}{\Delta^3} + \frac{1}{\Delta^2} \bigg[ \frac{4 \lambda (r-M)K}{\Delta} \\
&+& 4 \omega r \lambda + 12 \omega M \bigg] \bigg\} \,, \\
\nn b_2 &=& 4 \bigg\{\frac{8 K^3}{\Delta^2} + \frac{4 K}{\Delta} \bigg[ \frac{2(M^2-a^2)}{\Delta} - \lambda \bigg] \\
&-& \frac{8 \omega}{\Delta} (Mr-a^2) \bigg\}\,.
\eeq
\vspace{0.1cm}
The radial functions $\alpha$ and $\beta$ which define the Detweiler's function in Eq.~\eqref{DetweilerX} are
\comment{
\beq
\alpha &=& \frac{\kappa \mathfrak{a} \Delta^2 + |\kappa|^2}{\sqrt{2}|\kappa| \left(a_1 \Delta^2 + \rm{Re} \kappa\right)^{1/2}} \,, \\
\beta &=& \frac{i \kappa b_2 \Delta^2}{\sqrt{2} |\kappa| \left(a_1 \Delta^2 + \rm{Re} \kappa\right)^{1/2}} \,,
\eeq
}
where
\comment{
\beq
\nn \kappa &=& 4 \left[\lambda^2 (\lambda+2)^2 + 144a^2 \omega^2 (m-a\omega)^2 \right.\\
\nn &-& \left. a^2 \omega^2 (40\lambda^2-48\lambda)+a\omega m (40\lambda^2+48\lambda)\right]^{1/2} \\ 
&+& 48 i \omega M \,, \\
\nn \rm{Re} \kappa &=& 4 \left[\lambda^2 (\lambda+2)^2 + 144a^2 \omega^2 (m-a\omega)^2 \right.\\
\nn &-& \left. a^2 \omega^2 (40\lambda^2-48\lambda)+a\omega m (40\lambda^2+48\lambda)\right]^{1/2} \,, \\ \\
\nn |\kappa| &=& \big\{16 \left[\lambda^2 (\lambda+2)^2 + 144a^2 \omega^2 (m-a\omega)^2 \right. \\
\nn &-& \left. a^2 \omega^2 (40\lambda^2-48\lambda)+a\omega m (40\lambda^2+48\lambda)\right]\\ 
&+& \left(48 \omega M\right)^2 \big\}^{1/2}  \,.
\eeq
}
With this choice of parameters, $\alpha$ and $\beta$ satisfy the following relation
\comment{
\be
\alpha^2 - \alpha' \beta \Delta^{s+1} + \alpha \beta' \Delta^{s+1} - \beta^2 \Delta^{2s+1} V_s = \kappa \,,
\ee
}
which guarantees that the Detweiler's function defined in Eq.~\eqref{DetweilerX} satisfies 
Eq.~\eqref{final}. \comment{Furthermore, the conserved flux of energy is the same if 
computed by two independent solutions of Teukolsky's equation [Eq.~\eqref{wave_eq}] or two 
independent solutions of Detweiler's equation [Eq.~\eqref{final}] ~\cite{10.2307/79029}. This is an important 
consistency check since the energy flux is a measurable quantity and cannot depend on the 
trasformation of the perturbation variable.}

\comment{Equation~\eqref{pot_detweiler} gives the following potential
\be
V(r,\omega) = \frac{-K^2 + \Delta \lambda}{(r^2+a^2)^2} + \frac{\Delta (b_2 p' \Delta)'}{(r^2+a^2)^2 b_2 p} + G^2 + \frac{dG}{dr_*} \,, \label{pot_detw_final}
\ee
where
\be
p \equiv |\kappa| \left[ 2 \left( a_1 \Delta^2 + \rm{Re} \kappa \right)\right]^{-1/2} \,.
\ee
}
The effective potential \eqref{pot_detw_final} is purely real and has the following asymptotic forms: 
$V(r\to+\infty,\omega) \to - \omega^2$ and $V(r\to r_+,\omega) \to - \tilde{\omega}^2$.

\bibliographystyle{apsrev4}
\bibliography{Ref}

\end{document}